\documentclass[fleqn,usenatbib]{mnras}

\usepackage{newtxtext,newtxmath}

\usepackage[T1]{fontenc}

\DeclareRobustCommand{\VAN}[3]{#2}
\let\VANthebibliography\thebibliography
\def\thebibliography{\DeclareRobustCommand{\VAN}[3]{##3}\VANthebibliography}

\usepackage{graphicx}	
\usepackage{amsmath}	
\usepackage{multirow}
\usepackage{physics}
\usepackage{soul} 

\usepackage{wasysym}

\newcommand{\mps}{\;m\;s$^{-1}$\,}	
\newcommand{\kmps}{\;km\;s$^{-1}$\,}	
\newcommand{\bl}{B_\ell}	
\newcommand{\prot}{$P_\text{rot}$\;} 
\newcommand{\roundv}{\upsilon}       
\newcommand{\vsin}{\upsilon_\mathrm{eq} \sin i}       

\newcommand{\vrad}{$\roundv_\text{rad}$\;} 

\title[Magnetic field, accretion and planet of GM Aur]{Magnetic field, magnetospheric accretion and candidate planet of the young star GM Aurigae observed with SPIRou}

\author[B. Zaire et al.]{B. Zaire$^{1,2}$\thanks{E-mail: zaire@fisica.ufmg.br},
J-F Donati$^{2}$,
S. P. Alencar$^{1}$,
J. Bouvier$^{3}$,
C. Moutou$^{2}$,
S. Bellotti$^{4}$,
A. Carmona$^{3}$,
P. Petit$^{2}$,
\newauthor
\'A. K\'osp\'al$^{5,6}$,
H. Shang$^{7}$,
K. Grankin$^{8}$,
C. Manara$^{9}$,
E. Alecian$^{3}$,
S.~P. Gregory$^{10}$,
P. Fouqué$^{2}$,
\newauthor
and the SLS consortium
\\
$^{1}$ Universidade Federal de Minas Gerais, Belo Horizonte, MG, 31270-901, Brazil\\
$^{2}$ IRAP, Université de Toulouse, CNRS / UMR 5277, CNES, UPS, 14 avenue E. Belin, Toulouse, F-31400 France\\
$^{3}$ Université Grenoble Alpes, CNRS, IPAG, 38000 Grenoble, France\\
$^{4}$ Leiden Observatory, Leiden University, PO Box 9513, 2300 RA Leiden, The Netherlands\\
$^{5}$ Konkoly Observatory, HUN-REN Research Centre for Astronomy and Earth Sciences, Konkoly-Thege Mikl\'os \'ut 15-17, 1121 Budapest, Hungary\\
$^{6}$ Institute of Physics and Astronomy, ELTE E\"otv\"os Lor\'and University, P\'azm\'any P\'eter s\'et\'any 1/A, 1117 Budapest, Hungary\\
$^{7}$ Institute of Astronomy and Astrophysics, Academia Sinica, 10617 Taipei, Taiwan\\
$^{8}$ Crimean Astrophysical Observatory, 298409 Nauchny, Republic of Crimea\\
$^{9}$ European Southern Observatory, Karl-Schwarzschild-Straße 2, D-85748, Garching bei München, Germany\\
$^{10}$ SUPA, School of Science and Engineering, University of Dundee, Nethergate, DD1 4HN, Dundee, UK
}

\date{Accepted XXX. Received YYY; in original form ZZZ}

\pubyear{2024}

\begin{document}
\label{firstpage}
\pagerange{\pageref{firstpage}--\pageref{lastpage}}
\maketitle

\begin{abstract}
This paper analyses spectropolarimetric observations of the classical T Tauri star (CTTS) GM~Aurigae collected with SPIRou, the near-infrared spectropolarimeter at the Canada–France–Hawaii Telescope, as part of the SLS and SPICE Large Programs. We report for the first time results on the large-scale magnetic field at the surface of GM~Aur using Zeeman Doppler imaging. Its large-scale magnetic field energy is almost entirely stored in an axisymmetric poloidal field, which places GM~Aur close to other CTTSs with similar internal structures. A dipole of about $730$\,G dominates the large-scale field topology, while higher-order harmonics account for less than 30 per-cent of the total magnetic energy. Overall, we find that the main difference between our three reconstructed maps (corresponding to sequential epochs) comes from the evolving tilt of the magnetic dipole, likely generated by non-stationary dynamo processes operating in this largely convective star rotating with a period of about $6$\;d. Finally, we report a $5.5\sigma$ detection of a signal in the activity-filtered radial velocity data of semi-amplitude $110\pm20$\mps at a period of $8.745\pm0.009$\;d. If attributed to a close-in planet in the inner accretion disc of GM~Aur, it would imply that this planet candidate has a minimum mass of $1.10 \pm 0.30\;M_\mathrm{Jup}$ and orbits at a distance of $0.082 \pm 0.002$\;au.
\end{abstract}




\section{Introduction}
T Tauri stars (TTSs) are late-type pre-main-sequence (PMS) stars that are precursors of solar-like stars. These objects can be classified as classical T Tauri stars (CTTSs) when they show evidence of ongoing accretion from a surrounding disc and, later, as weak-line T Tauri stars (WTTSs) when accretion ceases due to the depletion of the inner disc. The study of TTSs has been crucial in advancing our understanding of planet formation and evolution, as their circumstellar discs are believed to be natural birthplaces for planets.  

Stellar magnetic fields impact the accretion process in CTTSs through a mechanism known as magnetospheric accretion \citep[see review by][]{HHC16}. CTTSs indeed possess kG magnetic fields that can truncate the inner disc and channel material onto the stellar surface through accretion funnels \citep[e.g.,][]{BAB07,ABW12}. The star-disc magnetospheric interactions also affect angular momentum transport through accretion, winds, and jets, thereby impacting stellar evolution \citep{BMM14}. Simultaneously, TTSs undergo gravitational contraction during their PMS evolution, leading to rapid changes in their internal structure. Variations in internal structure and stellar rotation period are believed to play a critical role in the dynamo mechanism that allegedly amplifies the magnetic field in these stars \citep[e.g.,][]{SN01}. There is direct observational evidence of the relationship of surface field strengths, age and rotation \citep[e.g.,][]{VGJ14}. Various theoretical models of self-excited dynamos also demonstrate distinct magnetic field solutions when internal structure and rotation rates vary \citep[e.g.,][]{ZGK17,EVB17,GZS19,BSN22}. These results suggest that the convection zone depth, rotation period, and radial mass distribution can impact the dynamo action. Overall, it is clear that the study of T Tauri stars can help elucidate the magnetic field amplification process as the internal structure and rotation rate rapidly change during this evolutionary phase.

Only in the past two decades have studies of TTSs made it possible to directly reconstruct large-scale magnetic fields using Zeeman-Doppler imaging \citep[ZDI,][]{S89,DB97,DHJ06}. Several optical and near-infrared spectropolarimetric programs, such as the ‘Magnetic Protostars and Planets’ (MaPP), ‘Magnetic Topologies of Young Stars and the Survival of massive close-in Exoplanets’ (MaTYSSE), ‘History of the Magnetic Sun’  \citep[HMS,][]{FPB16}, and more recently, the ‘SPIRou Legacy Survey’ \citep[][]{DKM20} conducted at the Canada-France-Hawaii Telescope (CFHT), have contributed to mapping the large-scale magnetic fields of TTSs. These studies unveiled in particular a correlation between the complexity of the magnetic field and the internal structure of the star \citep{GDM12}, similar to that observed on main-sequence M dwarfs \citep{MDP10}. Stars with fully or largely convective interiors exhibit strong axisymmetric poloidal fields dominated by low-order spherical harmonics components such as the dipole and octupole. In contrast, TTSs with substantial radiative interiors display weaker and more complex fields, often featuring a significant toroidal component \citep{DGA13}. However, the sample of CTTSs stars with reconstructed large-scale magnetic fields remains limited given the complicated effects induced by accretion process, whose temporal variability occurs on various timescales. Only a few dozen CTTSs have been analyzed to date, such as  V2129~Oph \citep{DJG07,DBW11}; BP~Tau \citep{DJG08}; CV~Cha and CR~Cha \citep{HCJ09}; V2247~Oph \citep{DSB10a}; AA~Tau \citep{DSB10b}; TW~Hya \citep{DGA11,DCL24}; V4046~Sgr \citep{DGM11};  GQ~Lup \citep{DGA12}; DN Tau \citep{DGA13};  LkCa~15 \citep{DBA19}; CI~Tau \citep{DBA20}; HQ~Tau \citep{PBA20}; V2062~Oph \citep{BAA20}; V807~Tau \citep{PBA21}; DQ~Tau \citep{PKK23}; DK~Tau \citep{NMR21proc, NMF23}; S~CrA~N \citep{NAP23}.  

In a complementary approach, researchers have also turned to global dynamo simulations to gain insights into the physical mechanisms governing the geometry of dynamo-generated magnetic fields \citep[e.g.,][]{C10,GDW12,YCM15}. These early simulations, albeit conducted with values of viscosity and turbulence that are not representative of actual stellar conditions, have managed somehow to reproduce the tentative link between interior and large-scale topology suggested by observations \citep{GDM12}. They have shown the interplay between magnetic complexity and Rossby number, which represents the ratio between the stellar rotation period and the typical time-scale of a convective cell, thus encompassing both the influence of rotation and internal structure. However, the generalization of these findings to more realistic stellar parameters remains a topic of ongoing discussion \citep[e.g.,][]{RPD15,ZJG22,BSN22} as it is their extension to CTTSs as accretion is not considered in any of these simulations. Increasing the sample of CTTSs with reconstructed large-scale magnetic fields is therefore essential to impose more constraints on dynamo models and deepen our understanding of the physical mechanisms that drive the generation and evolution of magnetic fields in TTSs.

In this paper, we present spectropolarimetric observations collected with SPIRou, the near-infrared (nIR) spectropolarimeter installed at CFHT \citep{DKM20}.  With these observations, we can investigate the large-scale magnetic field and the magnetospheric accretion process of the CTTS GM~Aurigae.  We provide a brief overview of the evolutionary status of GM~Aur in Sec.~\ref{sec:evolution} and describe our observations in Sec.~\ref{sec:obsdata}. The investigation of longitudinal magnetic fields from Zeeman signatures in circularly-polarized line profiles is presented in Sec.~\ref{sec:GPbl}, while in Sec.~\ref{sec:zdi} we use ZDI to reconstruct the brightness distribution and large-scale magnetic field topology at the surface of GM~Aur. In Sec.~\ref{sec:GPrv}, we use the photospheric absorption lines to inspect radial velocity (RV) fluctuations induced by activity. Our results are discussed and summarized in Sec.~\ref{sec:discussion}.

\section{The star-disc system GM~Aurigae} \label{sec:evolution}
The PMS star GM~Aurigae is a young solar analogue located at a distance of $155.0^{+1.4}_{-2.0}$\,pc \citep{GBV21} in the Taurus-Aurigae star formation region \citep{BG06}.  The star, classified as K6 \citep{HH14}, hosts a circumstellar disc that has been extensively studied to look for the presence of forming planets.

The spectral energy distribution (SED) of GM~Aur shows excess infrared emission with a deficit in the near- to mid-infrared \citep{SSE89}. SED modelling shows that GM~Aur is surrounded by a disc with a large dust cavity \citep{RWA03,CDW05,HSG16}. Spatially resolved (sub)millimetre observations confirm the existence of a transitional disc inclined at $52.77{\degr}\pm0.05{\degr}$ with respect to the line of sight with an inner dusty cavity radius of 30–40\;au \citep{MER18,FVM20,HAD20}. Yet, a consensus has not been reached on whether the dust-depleted cavity was carved by undetected planet(s), by photoevaporation, MHD disk winds or a combination of all \citep[see, e.g.][]{ITF23}. Further, an inner disc of gas inclined by ${68{\degr}}^{+17{\degr}}_{-28{\degr}}$ has been detected within the dust cavity using observations from VLTI/GRAVITY (\citealt{BBP22}, but see also \citealt{SBB09,HAD20,BBL21}). 
Although less precise than the transition disc inclination, these measurements suggest that GM~Aur's inner and outer discs can be considered aligned as argued by \citet{BBP22}. Furthermore, the inner disc presumably feeds the accretion process resulting in accretion rates ranging from $0.2\times 10^{-8}$ to $2.15 \times 10^{-8}$\,M$_{\odot}\mathrm{yr}^{-1}$ \citep{ICH13,MTN14,IEC15,RE19,BSP23,WEK24}.
 
Long-term photometry of GM~Aur has shown low-level variability due to surface spots, and indicates a stellar rotation period in the range of $5.8$ to $6.1$ days \citep{PGS10,RER22}. More recently, \citet{BSP23} analyzed LCOGT light curves of GM~Aur that overlap with part of the observations studied in this paper. The authors found modulations consistent with a period of $6.04\pm0.15$\,d. Likewise, they used spectroscopy to demonstrate that the inverse P Cygni profiles of near-infrared accretion lines (HeI\,$1083$\,nm, Pa$\beta$, Br$\gamma$), whose sub-continuum absorption traces accretion funnels rotating into and out of view \citep[e.g.,][]{HHC94,MCH01,EFH06}, are consistent with the $6.04$\,d period obtained from photometry.

\begin{table}
	\centering
	\caption{Properties of GM~Aur.}
	\label{tab:StellarProperties}
	\begin{tabular}{llll}
		\hline
		Parameter  &  Value & Comments & Reference\\
		\hline
        Distance (pc) & $155.0^{+1.4}_{-2.0}$ &  & 1 \\
  \noalign{\smallskip}
		Age (Myr) & $\sim 1.5$ & from HR diagram & Sec.~\ref{sec:evolution}\\
  \noalign{\smallskip}
        Spectral type & K6 & & 2,3 \\
 \noalign{\smallskip}
		$M_\star$ ($M_{\odot}$)  &  $0.95\pm0.05$ & from HR diagram & Sec.~\ref{sec:evolution} \\
  \noalign{\smallskip}
		$T_\text{eff}$ (K) & $4287 \pm 35$ &  & 4 \\
    \noalign{\smallskip}
		$i$ (${\degr}$)  & $52.77 \pm 0.05$ & from outer disc & 5 \\  
  \noalign{\smallskip}
		$\vsin$ (km\,s$^{-1}$) & $13.5 \pm 0.2$ & from ZDI optimisation & Sec.~\ref{sec:zdi} \\
  \noalign{\smallskip}

        \prot (d)   & $6.04$ & used to phase data & 4 \\
  \noalign{\smallskip}
    	\prot (d)   & $6.03^{+0.03}_{-0.04}$ & from $\bl$ data & Sec.~\ref{sec:GPbl} \\
  \noalign{\smallskip}
  		\prot (d)   & $5.99\pm0.01$ & from RV data & Sec.~\ref{sec:GPrv} \\
  \noalign{\smallskip}
		$R_\star\sin i$ ($R_{\odot}$)  &  $1.61\pm 0.05$ & from $\vsin$ and \prot & Sec.~\ref{sec:evolution} \\
  \noalign{\smallskip}
		$R_\star$ ($R_{\odot}$)  &  $2.02\pm 0.06$ & from $\vsin$, \prot and $i$ & Sec.~\ref{sec:evolution} \\
  \noalign{\smallskip}
		\vrad (km\,s$^{-1}$)  & $14.9 \pm 0.3$ &  & Sec.~\ref{sec:GPrv} \\
  \noalign{\smallskip}
        $L_\star$ ($L_{\odot}$) & $1.25 \pm 0.08$ & from $T_\text{eff}$ and $R_\star$ & Sec.~\ref{sec:evolution} \\
\noalign{\smallskip}
		$\log g$ (dex) & $3.80\pm 0.03$ & from $M_\star$ and $R_\star$ & \\
  \noalign{\smallskip}
		$\xi$ (km\,s$^{-1}$) & 1.7 &  & 6 \\
  \noalign{\smallskip}
  		$r_\mathrm{cor}$ ($R_\star$) & $6.8\pm0.2$ &  & Sec.~\ref{sec:discussion} \\
\noalign{\smallskip}
      	$r_\mathrm{m}$ ($R_\star$) & $4.1\pm1.0$ &  & Sec.~\ref{sec:discussion} \\
		\hline
	\end{tabular}
	{\bf Note.} (1) \citet{GBV21}; (2) \citet{HH14}; (3) \citet{L18}; (4) \citet{BSP23}; (5) \citet{MER18}; (6) \citet{DBR11}.
\end{table}

By combining the line-of-sight-projected equatorial rotation velocity inferred from our spectropolarimetric analysis ($\vsin = 13.5 \pm 0.2$\kmps, Sec.~\ref{sec:zdi_maps}), rotation period ($P_\mathrm{rot} = 6.04 \pm 0.15$\,d, Eq.~\ref{eq:ephemeris}), effective temperature \citep[$T_\mathrm{eff} = 4287 \pm 35$\,K  estimated with spectral synthesis,][]{BSP23}, and assuming that the rotation axis of the star coincides with that of the disc ($i=52.77{\degr}\pm 0.05{\degr}$), we get a stellar radius of $R_\star = 2.02\pm0.06$\,R$_{\odot}$. With this radius estimate, we infer a bolometric luminosity equal to $1.25\pm0.08\,L_{\odot}$. These radius and luminosity estimates are both higher than those obtained previously by \citet{BSP23}, derived from the median J magnitude (J=$9.42$), and respectively equal to $1.7\pm0.2$\,R$_{\odot}$ and $0.9 \pm 0.2$\,L$_{\odot}$. The difference can presumably be attributed to cool spots at the surface reducing the luminosity \citep{GSH17}.  Table~\ref{tab:StellarProperties} summarizes the stellar properties of GM~Aur used in our paper. 

\begin{figure}
	\includegraphics[width=\columnwidth]{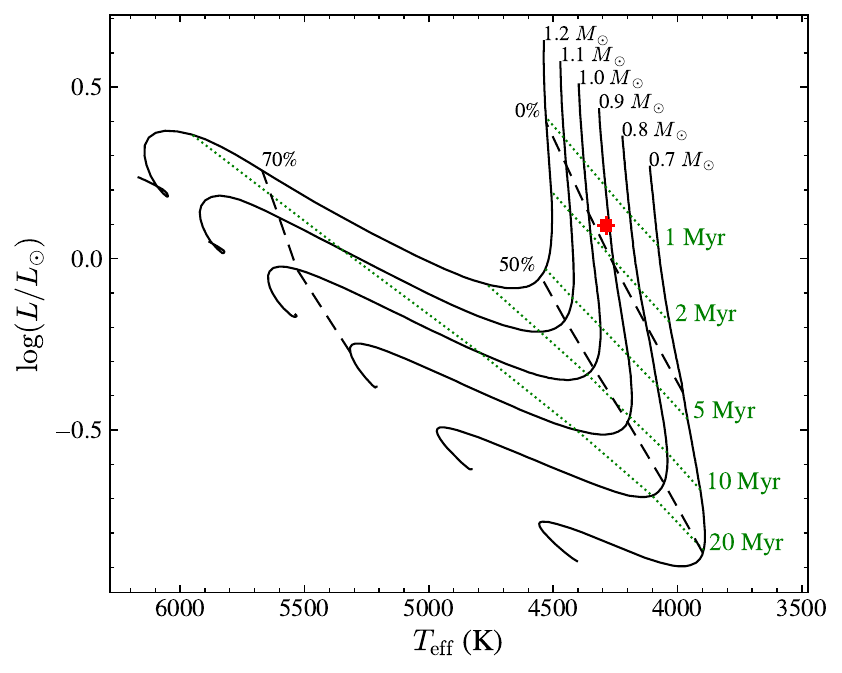}
    \caption{Location of GM~Aur (red square) in the Hertzsprung–Russell diagram constructed from \citet{BHA15} evolutionary models. Pre-main sequence evolutionary tracks are for stellar masses of $0.7$, $0.8$, $0.9$, $1.0$, $1.1$, and $1.2 \;M_{\odot}$ (black solid lines). Isochrones are for ages of $1$, $2$, $5$, $10$, and $20$\;Myr (dotted green lines). Dashed black lines mark the internal structure change, corresponding to the location where the radiative core starts forming and when its radius reaches $50\%$ and $70\%$ of the stellar radius.}
    \label{fig:hrd}
\end{figure}

Figure~\ref{fig:hrd} illustrates the evolutionary stage of GM~Aur. Its location in the Hertzsprung–Russell (HR) diagram, compared to the PMS stellar evolutionary tracks of \citet{BHA15}, indicates an age of $\sim 1.5$\;Myr and a stellar mass of $0.95 \pm 0.05$\,M$_{\odot}$. This mass range is consistent, within a 3$\sigma$ level, with the independent measurement made by \citet{SGD17}, who analyzed CO emission lines from GM~Aur's circumstellar disc to infer a dynamical mass of $1.14 \pm 0.02$\;M$_{\odot}$. Regarding its internal structure, GM~Aur lies near the fully convective limit. Evolutionary models suggest the star is fully convective for a luminosity of $1.25\pm0.08\,L_{\odot}$ (red square in Fig.~\ref{fig:hrd}), while for a luminosity of $0.9 \pm 0.2$\,L$_{\odot}$ it may possess a small radiative core, estimated to be up to $0.2\pm0.1 R_\star$ and comprising no more than $15\%$ of the stellar mass.

In the following, we consider that GM~Aur is a star of mass $0.95\,M_{\odot}$ and of radius $2.02\pm0.06$\,R$_{\odot}$. We adopt the ephemeris of \citet{BSP23} to compute the rotation cycle $E$:
\begin{equation} \label{eq:ephemeris}
    \text{BJD} \, = \, 2459460.80 + 6.04 E.
\end{equation}
It is worth noting that the period determination of \citet{BSP23} is based on LCOGT light curves that were collected simultaneously with the first two SPIRou runs analyzed in this paper. Therefore, using the same ephemeris eases the comparison between both studies.

\section{Spectropolarimetric observations} \label{sec:obsdata}
Spectropolarimetric observations of GM~Aur were collected from September 2021 to January 2023 using SPIRou at CFHT. These observations were part of the CFHT Large Programs called ‘SPIRou Legacy Survey’\footnote{\url{http://spirou.irap.omp.eu/Observations/The-SPIRou-Legacy-Survey}} (SLS, PI: Jean-Fran\c{c}ois Donati) and SPICE (PI: Jean-Fran\c{c}ois Donati). Studying the impact of magnetic fields on star/planet formation is one of the main science goals of SLS and SPICE.

SPIRou operates in the nIR domain, covering wavelengths from $950$\;nm to $2500$\;nm at a resolving power $\mathcal{R} \sim 70000$ \citep{DKM20}. Raw frames were processed with the SPIRou reduction pipeline \verb|APERO| \citep{CAD22}, which produces wavelength calibrated continuum normalized spectra corrected of telluric absorption. Data products are thus used to extract unpolarized (Stokes $I$) and polarized (Stokes $V$) spectra by the \texttt{spirou-polarimetry}\footnote{\url{https://github.com/edermartioli/spirou-polarimetry}} package. The package follows the prescription of \citet{DSC97} for optimal extraction of the polarized spectra. It combines four consecutive sub-exposures obtained in different orientations of the polarimeter quarter-wave Fresnel rhombs, chosen to minimize spurious polarisation signatures \citep{CAD22}. For GM Aur,  49 polarized spectra were derived from the sequences of 4 sub-exposures (each lasting $552$\;s). The journal of observations is summarised in Table~\ref{tab:results}.

\subsection{Least Squares Deconvolution}  \label{sec:lsd} 
Least-Squares Deconvolution \citep[LSD][]{DSC97} is applied to all observations to generate average Zeeman signatures with an enhanced signal-to-noise ratio. LSD Stokes $I$ and $V$ profiles are generated with the open-source software \texttt{LSDpy}\footnote{\texttt{LSDpy} is available at \url{https://github.com/folsomcp/LSDpy}.}, which uses a list of photospheric absorption lines and user-defined normalisation parameters to determine average pseudo-profiles. We use the VALD3 database\footnote{VALD3 is publicly available at \url{http://vald.astro.uu.se}.} \citep{PKR95} to build an atomic absorption line mask for GM~Aur that covers the spectral domain of SPIRou. A MARCS model atmosphere \citep{GEE08} with an effective temperature of $4250$\;K, a logarithmic surface gravity of $4.0$, and microturbulence $\xi = 1.7$\kmps is considered, all of which agree with the stellar parameters of GM~Aur (see Table~\ref{tab:StellarProperties}). In particular, the logarithmic surface gravity considered roughly agrees with estimates based on the stellar mass and radius ($\log g = 3.80\pm 0.03$) and previously reported in the literature \citep[$\log g = 3.91^{0.06}_{-0.08}$,][]{FCR22}. As an additional condition, we remove from the line mask those lines whose absorption depth is lower than $5\%$ of the continuum (before any kind of broadening). Altogether, our line mask contains a total of 2476 spectral features. Normalisation parameters needed to compute LSD profiles are the central wavelength  $\lambda_0$, the line depth $d$, and the effective Landé factor $g_\mathrm{eff}$. Based on the median values of our line mask, we set $\lambda_0 = 1700$\;nm, $d = 0.19$, and $g_\mathrm{eff} = 1.25$. Stokes $V$ LSD profiles show average noise levels with respect to the unpolarized continuum of  $1.22\times 10^{-4}$ to $2.04 \times 10^{-4}$ (median of $1.49 \times 10^{-4}$). The null polarisation check $N$ \citep[as defined in][]{DSC97} is consistent with 0 at all times (see Appendix~\ref{appendix:unveil}), confirming that the polarimeter performs as expected.

We also considered an alternative molecular line mask based on the magnetic-insensitive transitions from the CO molecule. The mask includes the CO bandhead lines in the wavelength range from $2200$\,nm to $2400$\,nm, displaying a mean wavelength of $2354$\;nm. The LSD profiles obtained from the CO bandhead lines are considered only in Sec.~\ref{sec:GPrv}. Hereafter, unless we explicitly mentioned otherwise, whenever Stokes $I$ and $V$ LSD profiles are mentioned they refer to profiles from atomic lines.

We double-checked that both our polarized spectra and LSD profiles yielded similar results to those independently obtained with the alternate reduction package \texttt{Libre-ESpRIT} and LSD code outlined in \citet{DLC23}.

\subsection{Correcting for the veiling} \label{sec:veiling}

Photospheric lines of CTTSs are veiled by excess continuum emission \citep{HEG95,HHC16} that has different origins depending on the spectral domain. While accretion shocks are often the main source of optical veiling, the warm inner disc is thought to emit most of the radiation that veils the nIR spectral domain \citep[e.g.,][]{JKV01}. For GM~Aur, \citet{SBA23} found a veiling wavelength dependence in which the Y-band veiling was consistent with 0, while in the J, H and K-bands it was $0.13$, $0.11$, $0.31$, respectively. Moreover, \citet{BSP23} showed that the veiling contribution in the JHK bands varies over time in a periodic way, featuring a maximum value around the rotational phase 0. As our paper aims to investigate line profile variations caused by features at the stellar photosphere, we must first remove the veiling variability introduced on Stokes $I$ and $V$ LSD profiles.

Fluctuations in the equivalent width (EW) of Stokes $I$ LSD profiles are primarily affected by veiling but can also probe temperature variations at the surface of the star \citep[e.g., LkCa~4,][]{GSH17,FDG23}. Assuming that the effect of veiling largely dominates, one can mitigate it by scaling Stokes $I$ and $V$ LSD profiles to ensure a constant EW in the data set. To suppress the veiling variability, we scale the LSD profiles to ensure a constant EW of 1.11\kmps in the data set. The results of the EW measurements before applying the scaling procedure are presented in Table~\ref{tab:results}. Equivalent widths ranged from $0.84$ to $1.11$\kmps (Table~\ref{tab:results}), with a median value of 1.02km\;s$^{-1}$. See Appendix~\ref{appendix:unveil} for further discussion about the calculus of unveiled Stokes $I$ and $V$ LSD profiles. 
 
\section{Longitudinal magnetic field} \label{sec:GPbl}
We used the \verb|specpolFlow| package\footnote{The \texttt{specpolFlow} package is publicly available at \url{https://github.com/folsomcp/specpolFlow}.} to estimate the longitudinal magnetic field ($\bl$ in Gauss) as the first-order moment of Stokes $V$ LSD profiles \citep{DSC97}:
\begin{equation} \label{eq:blong}
    \bl = - 2.14 \times 10^{11} \frac{\int \roundv \cdot V(\roundv) \mathrm{d}\roundv}{\lambda_0 \cdot g_\mathrm{eff} \cdot c \cdot \int [1 - I(\roundv)] \mathrm{d}\roundv},
\end{equation}
where $c$ is the speed of the light in \kmps, $\roundv$ is the Doppler velocity in the stellar rest frame (also in \kmps), and $\lambda_0$ (in nm) and $g_\mathrm{eff}$ are the wavelength and magnetic sensitivity of the average LSD profile, i.e. the normalisation quantities used to compute Stokes $I$ and $V$ LSD profiles (Sec.~\ref{sec:lsd}). We adopted the \vrad estimation of $14.94$\kmps (see Sec.\ref{sec:GPrv}) to place Stokes $I$ and $V$ LSD profiles in the stellar rest frame. Table~\ref{tab:results} presents $\bl$ measurements obtained after evaluating the integral of Eq.~\ref{eq:blong} in a velocity window of $\pm 35$\;km\;s$^{-1}$. The values of $\bl$ ranged from 
$-158$\;G to $-30$\;G, with a median value of $-107$\;G and a standard deviation of $32$\;G.

\begin{figure*}
	\includegraphics[width=\textwidth]{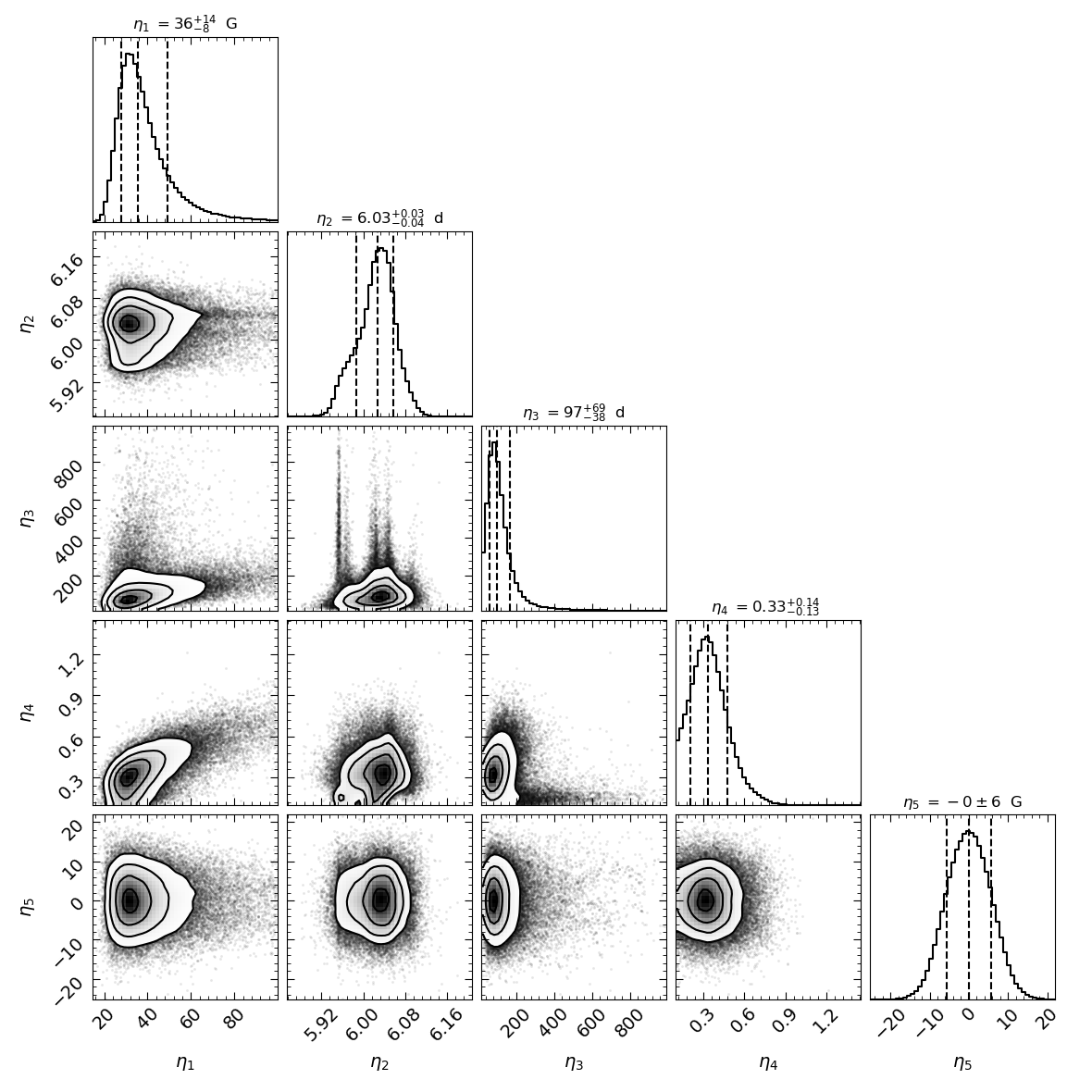}
    \caption{Posterior distributions obtained from an MCMC exploration over the hyperparameter space of the GP model of the longitudinal magnetic field evolution. The semi-amplitude $\eta_1$ and white noise $\eta_5$ are given in G, whereas the rotation period $\eta_2$ and cycle length $\eta_3$ are expressed in days, and the smoothing parameter $\eta_4$ is dimensionless. Vertical dashed lines in the histograms correspond to the 16th, 50th, and 84th percentiles of each distribution.}
    \label{fig:GP_corner}
\end{figure*}

We modeled the longitudinal magnetic field evolution using Gaussian Process (GP) regression. We selected for the GP model a quasi-periodic kernel added of a white noise kernel, whose ability to recover stellar rotation periods has been demonstrated in previous studies \citep{DYM17,YDH17,AMA18,PFD21}, and that is defined as 
\begin{equation}
    K(t,t') = \eta_1^2 \exp\left[-\frac{(t-t')^2}{2\eta_3^2} -\frac{1}{2\eta_4^2}\sin^2\left(\frac{\pi(t-t')}{\eta_2}\right)\right] + \eta_5^2 \delta(t-t').
\end{equation}
Five GP hyperparameters control the kernel. The hyperparameter $\eta_1$ describes the semi-amplitude of the GP, $\eta_2$ the rotation period of the star, $\eta_3$ the decay time (likely the typical timescale of evolution of active regions), $\eta_4$ the level of harmonic complexity allowed in the fit (the smaller, the more complex the fit), and $\eta_5$ the white noise (in case formal error bars are underestimated, e.g. as a result of intrinsic variability). To obtain the best statistical GP model of $\bl$ and its associated uncertainties, we maximized the log of the marginal likelihood function $\mathcal{L}_M$:
\begin{equation}
    2 \log \mathcal{L}_M = - n\log(2\pi) - \vb{y}^{T}(\vb{K} + \vb{\Sigma})^{-1}\vb{y} - \log\abs{\vb{K} + \vb{\Sigma}} ,
\end{equation}
where $n$ is the number of measurements, $\vb{y}$ is the $\bl$ data, $\vb{K}$ is the GP kernel covariance matrix, and $\vb{\Sigma}$ is a diagonal matrix of the data variance.

\begin{table}
    \caption[]{GP parameters obtained when fitting a quasi-periodic kernel to individual time series longitudinal field data of GM~Aur. The first and second columns indicate the hyperparameter and its associated variable name used throughout the paper. The third column shows the best value obtained in the GP fit, whereas the last column indicates the prior distribution adopted – for which $\mathcal{N}$ stands for Gaussian distribution and $\mathcal{U}$ for uniform distribution.}
    \label{tab:gpBlfitparams}
    \begin{tabular}{lccc}
        \hline
     Parameters & Name & Value & Prior \\
        \hline
     GP semi-amplitude  (G)    & $\eta_1$                  & $36^{+14}_{-8}$ & $\mathcal{U}(0,1000)$ \\
        \noalign{\smallskip}
     Rotation period (d)  & $\eta_2$ & $6.03^{+0.03}_{-0.04}$ & $\mathcal{N}(6.04,1)$ \\
        \noalign{\smallskip}
     Decay time scale (d) & $\eta_3$                  & $97^{+69}_{-38}$ & $\mathcal{U}(0,1000)$  \\
             \noalign{\smallskip}
     Smoothing factor  & $\eta_4$                  & $0.33^{+0.14}_{-0.13}$ & $\mathcal{U}(0.1,3)$ \\
        \noalign{\smallskip}
     White noise  (G)      & $\eta_5$                  & $0\pm6$  & $\mathcal{U}(-1000,1000)$ \\
        \noalign{\smallskip}
     \hline
    \end{tabular}
\end{table}

We conducted a Markov Chain Monte Carlo (MCMC) exploration over the GP hyperparameter space using the \verb|emcee| package \citep{FMH13}\footnote{The \texttt{emcee} python package is freely available at \url{https://github.com/dfm/emcee}.} to sample the posterior distribution. We adopted uniform prior probability distributions for the hyperparameters $\eta_1$, $\eta_3$, $\eta_4$, and $\eta_5$ (Table~\ref{tab:gpBlfitparams}). In contrast, we assumed a Gaussian prior for $\eta_2$, centered on $6.04$\;d \citep{BSP23}. We initialised the MCMC sampler with $32$ walkers. Chains ran until a convergence criterion of $230$ times the autocorrelation time $\tau$\footnote{Autocorrelation times are defined as the number of iterations necessary for the MCMC sample to become independent of previous draws for each GP hyperparameter \citep{GW10}. In this study, we define as $\tau$ the largest value among the autocorrelation times estimated for a chain.} was reached. The first $50\;\tau$ were discarded as burn-in. MCMC chains featured an auto-correlation time $\tau$ equal to $525$ steps.

\begin{figure*}
	\includegraphics[width=\textwidth]{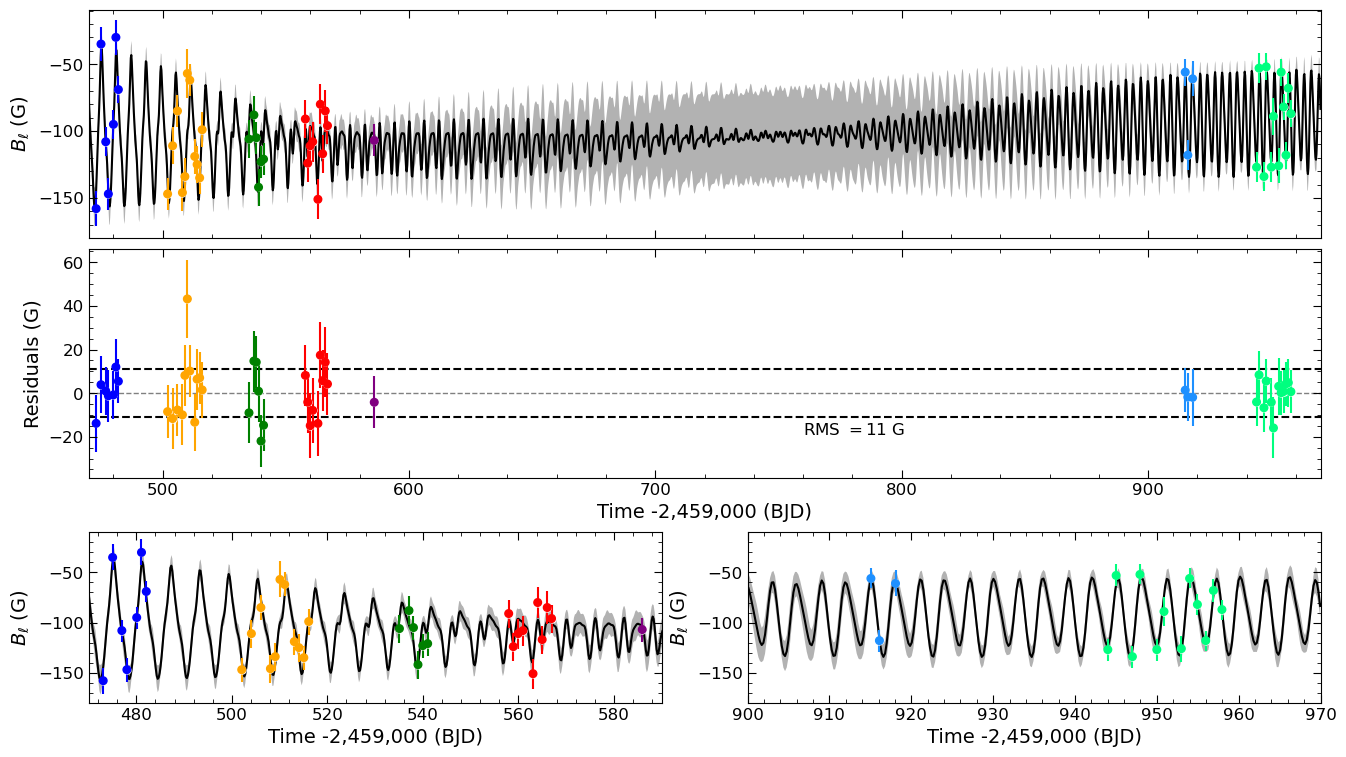}
 \caption{Gaussian process analysis of the longitudinal magnetic field of GM~Aur. The top panel illustrates longitudinal magnetic field measurements (colored circles) alongside the GP-prediction values (black line). 1$-\sigma$ error bars are depicted for both measurements and predictions. Colors represent SPIRou runs: September (blue), October (orange), November (green), and December 2021 (red), January and December 2022 (purple and light blue, respectively), and January 2023 (light green). The second panel shows residuals (observed minus modeled field). The last row offers detailed views of observations from October 2021 to January 2022 (left-hand panel) and December 2022 to January 2023 (right-hand panel).}
    \label{fig:GPbl}
\end{figure*}

Fig.~\ref{fig:GP_corner} shows the posterior distributions obtained from the MCMC exploration. We quote in Table~\ref{tab:gpBlfitparams} the median values of the posterior distributions, taken as representative of the best-fit solution. Figure~\ref{fig:GPbl} illustrates the best-fit GP model of $\bl$, which fits the data down to a reduced-$\chi^2$ of 0.7 and displays a root-mean-square (RMS) value of $11$\;G in the residuals. The rotation period of $6.03^{+0.03}_{-0.04}$\;d obtained from the analysis of $\bl$ agrees within uncertainties with the estimation in our companion study by \citet{BSP23}, who investigated the variability in the light curve of GM~Aur and found a periodicity of $6.04 \pm 0.15$\;d.

\section{Zeeman-Doppler imaging} \label{sec:zdi} 
In this section, we use the tomographic method used in a series of works \citep{BDR91,DB97,DHJ06} to reconstruct the large-scale magnetic field morphology of GM~Aur using time-series Stokes $I$ (unpolarized) and $V$ (polarized) LSD profiles. 

\subsection{Zeeman-Doppler imaging process} \label{sec:modeldescription}  
Zeeman-Doppler Imaging is an inversion method that maps brightness and magnetic inhomogeneities in the stellar photosphere from phase-resolved sets of Stokes $I$ and $V$ LSD profiles. The basic assumption behind  ZDI is that these distortions/signatures mainly vary in time as a result of rotational modulation. To achieve this, ZDI divides the surface of the star into a grid of $N$ cells (here, set to $4000$ cells). These cells are independent of each other for brightness imaging. In contrast, magnetic images are described as a spherical-harmonics (SH) expansion (as described in \citealt{DHJ06} and using the coefficient modification detailed in \citealt{LD22}). The optimisation procedure is initialised from a featureless brightness and magnetic image. Then, ZDI uses Maximum Entropy principles to iteratively search for images that maximise the entropy, defined as the amount of information in the image, whilst aiming for a data fit at a given value of reduced-$\chi^2$ \citep[see][for a description of the maximum entropy regularization procedure]{SB84}. Like in previous studies of T Tauri stars \citep{DJG08}, we employ a weighting scheme in the image entropy calculation that favours even SH modes when reconstructing magnetic maps.

To compute the $\chi^2$ statistics, ZDI calculates the disc-integrated Stokes $I$ and $V$ profiles using modelled magnetic field and brightness images obtained at each iteration\footnote{Note that an alternative is to impose the brightness image in the modeling process (see Sec.~\ref{sec:zdi_maps}).}. Following the definition of \citet{MDP08}, we write the synthetic Stokes profiles in terms of the filling factors $f_I$ and $f_V$:
\begin{equation}
    I = f_I \cdot I_M + (1 - f_I) \cdot I_Q,
\end{equation}
and
\begin{equation}
    V = f_V \cdot V_M,
\end{equation}
where $f_I$ represents the typical fractional area of a cell that is filled with small-scale magnetic fields and $f_V$ accounts for the typical fractional area filled with large-scale magnetic fields capable of producing net circularly polarized signatures \citep[see][for visual examples of the impact of filling factors on the surface magnetic field]{K21}. Under this formulation, $I_Q$ represents the local absorption line profile obtained from non-magnetic regions, whereas $I_M$ and $V_M$ correspond to local Stokes $I$ and $V$ profiles from magnetic regions. $I_M$ and $V_M$ are computed using Unno-Rachkovsky's solution to the radiative transfer equation \citep[][Chapter 9.8]{LDL04}. 

The model parameters adopted for GM~Aur follow previous nIR studies \citep[e.g.,][]{DFC24}. We set the filling factors to $f_I=0.8$ and $f_V = 0.4$, meaning that $20\%$ of the stellar surface is non-magnetic. Those filling factors are similar to what was suggested by Zeeman broadening studies of the CTTS CI Tau \citep{SJM20} or the young M dwarfs AD Leo \citep{BML23} and AU Mic \citep{DCF23}. For the synthetic line model, we use the central wavelength of $1700$\,nm, Landé factor of $1.25$, and Doppler width of 3.5\kmps \citep[similar to][]{FDC23,DCF23}. Moreover, we truncate the SH expansion of the magnetic field components at modes $\ell=10$. Given GM~Aur's value of $\vsin$, SH coefficients with $\ell>10$ should not carry relevant information when modelling the stellar surface \citep[e.g.,][]{MDP08,FDJ12,FPB16}.
Finally, we assume $i = 53\degr$ in accordance with literature values (Table~\ref{tab:StellarProperties}).

One caveat of our ZDI model is that it does not handle intrinsic variability beyond differential rotation in its current version \citep[e.g.,][]{DYM17,FDK21,ZDK22} as it assumes static brightness and magnetic field maps throughout the observed window \citep[see a recent attempt to go beyond these limitations and recover the temporal evolution of large-scale magnetic field maps in][]{FD22}. Considering the rapid evolution of GM Aur's magnetic field within a short timescale (approximately $97$\;d, as discussed in Sec.~\ref{sec:GPbl}), we separated the SPIRou spectropolarimetric observations, spanning over 485 days, into three distinct datasets for independent analysis. These datasets comprise observations: $\#1$ from September and October 2021, $\#2$ from November 2021 to January 2022, and $\#3$ from December 2022 to January 2023.

\subsection{ZDI surface maps of GM Aurigae} \label{sec:zdi_maps}
We first attempted to model brightness maps simultaneously with magnetic fields. The reconstructed ZDI images revealed only low-level brightness inhomogeneities (compared to the unspotted photosphere) that have a small impact on the Stokes $I$ LSD profiles, suggesting that the profile broadening is mostly of magnetic origin. Given that, we chose to fit the LSD Stokes $I$ and $V$ profiles of GM~Aur using a simple ZDI model that imposes a constant (featureless) brightness map and only takes magnetic fields into account. By exploring the principles of maximum entropy, we find that $\vsin = 13.5\pm0.2$\kmps provides the best fit for the three data sets. This value is similar to previous infrared estimates of $13.7\pm1.7$\kmps using IGRINS spectra \citep{NJL21}. 

\begin{figure*}
	\includegraphics[width=0.82\textwidth]{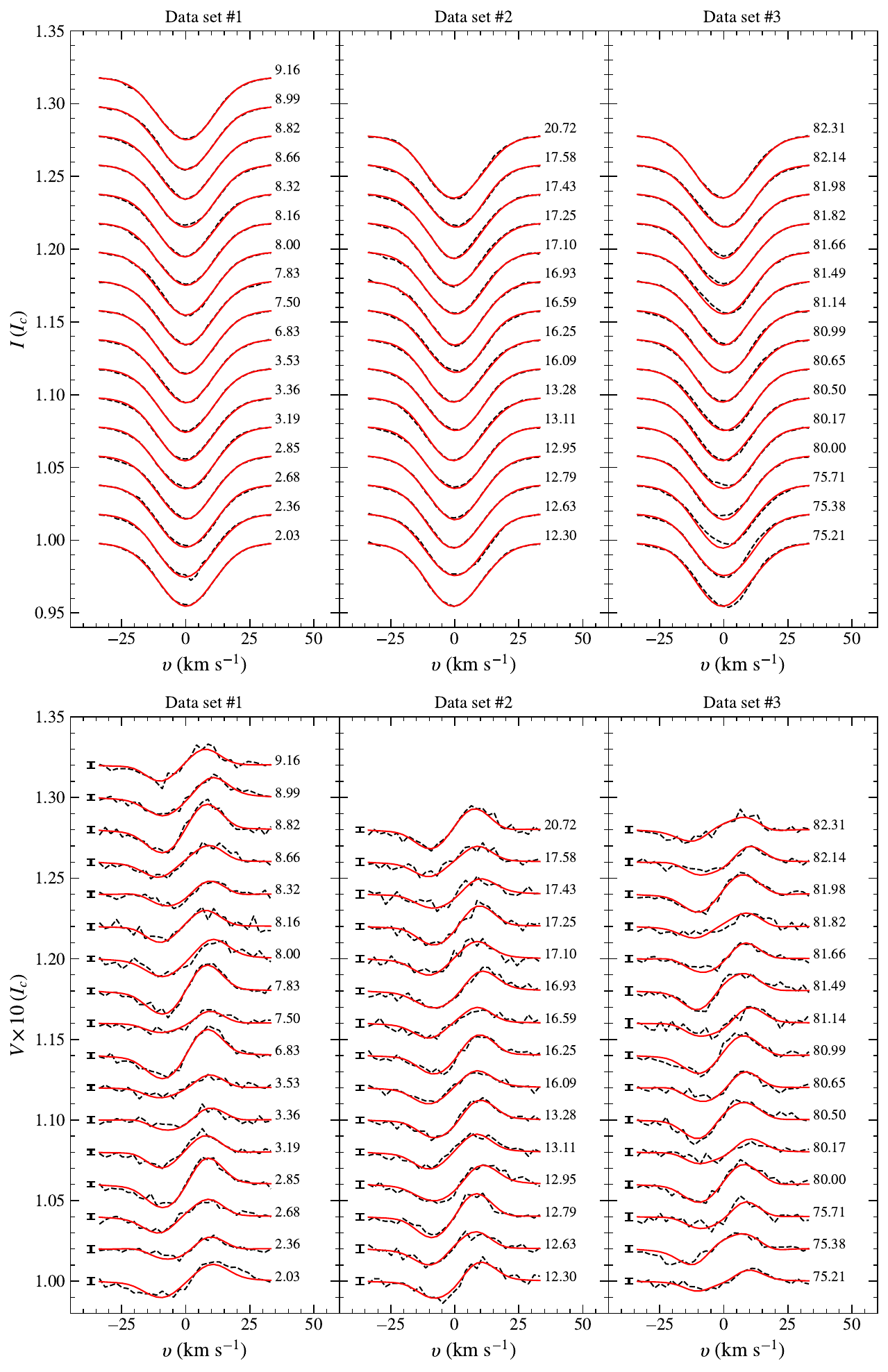}
    \caption{Maximum entropy fit to the Stokes $I$ (first row) and $V$ (second row) LSD observations (dashed black lines). Columns show the ZDI fit (red lines) to the data sets $\#1$, $\#2$, and $\#3$, respectively. Stokes $I$ and $V$ profiles have been vertically shifted for display purposes. Rotation cycles and $1\sigma$ error bars (only for the Stokes $V$ profiles) are shown next to each profile.}
    \label{fig:bestentropyfit}
\end{figure*}

The ZDI maximum entropy fit to the Stokes $I$ and $V$ LSD profiles achieved for the data sets $\#1$, $\#2$, and $\#3$ are displayed in Fig.~\ref{fig:bestentropyfit}. To obtain these ZDI models, we had to increase the error bars of the Zeeman signatures by a factor of 1.3 in order to account for intrinsic variability. This procedure allowed us to fit  Stokes $I$ and $V$ LSD profiles down to a reduced $\chi^2$ equal to $1$ (the same factor was applied to all three data sets). 

\begin{figure*}
	\includegraphics[width=\textwidth]{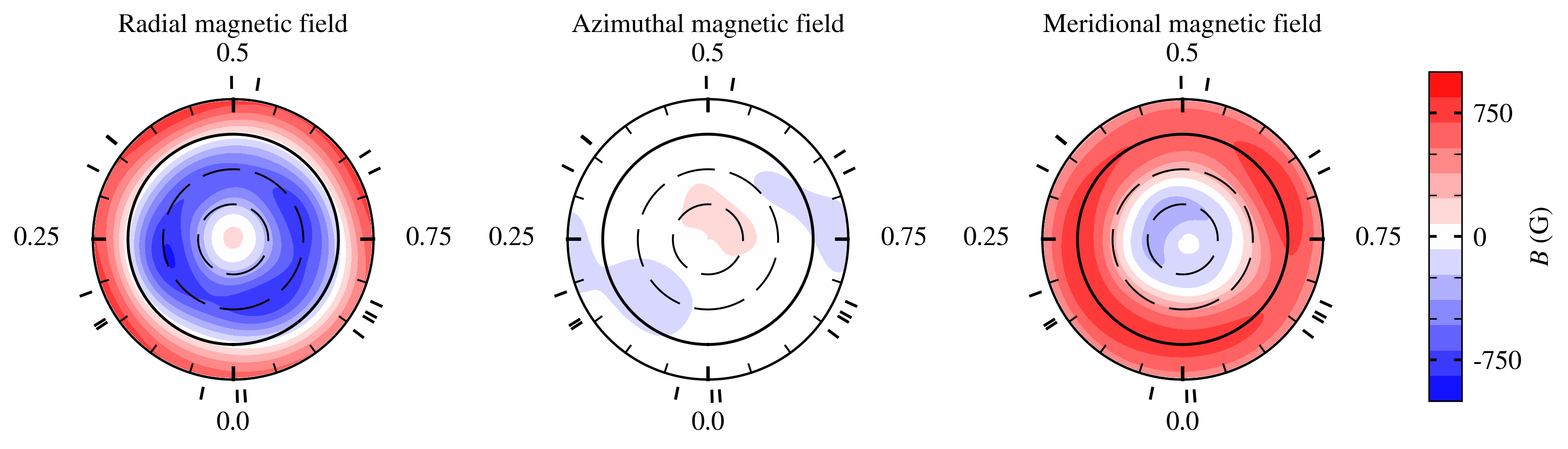}
	\includegraphics[width=\textwidth]{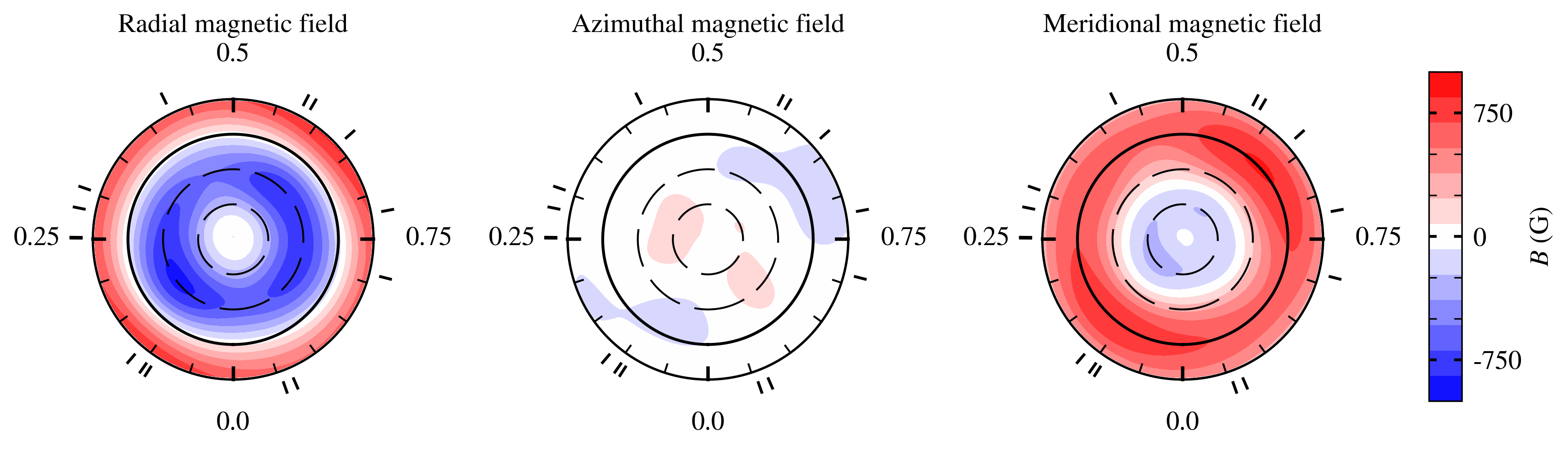}
	\includegraphics[width=\textwidth]{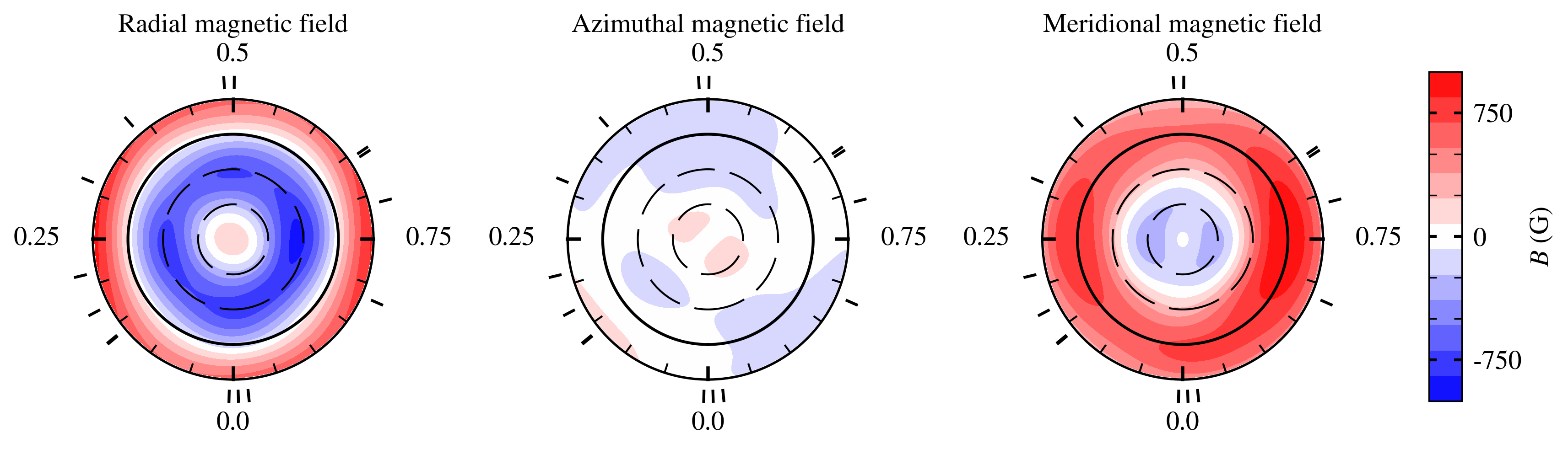}
    \caption{Magnetic field maps of GM~Aur reconstructed using data set $\#1$ (top panels), data set $\#2$ (middle panels), and data set $\#3$ (bottom panels). Maps are shown in flattened polar projection in which concentric circles represent $30{\degr}$ steps in latitude. Columns show respectively the radial, azimuthal, and meridional components of the magnetic field. Magnetic fields are given in units of Gauss with positive values represented in red shades and negative values in blue.}
    \label{fig:magmaps}
\end{figure*}

Figure~\ref{fig:magmaps} presents the surface maps reconstructed for GM~Aur. From left to right, columns display the surface distribution of the radial, azimuthal, and meridional magnetic field components. Overall, the large-scale magnetic topology is quite similar in all epochs. It consists of a dominant axisymmetric poloidal field that adds up to about $98\%$ of the total energy. Most of the differences between the maps are due to a minor change in the dipole field strength and orientation. The dipole that is tilted by $15{\degr}$ towards rotation phase $0.91$ in the first epoch slightly changes its obliquity to $10{\degr}$ in the last two epochs, facing the rotation phase $0.95$ in the second epoch and then phase $0.81$ in the last epoch. In all three epochs, the negative pole of the dipolar field component is in the northern hemisphere. As evidenced by the radial magnetic field maps, the octupolar field has the opposite polarity of the dipole field at the polar cap, resulting in nearly zero magnetic field strengths at colatitudes lower than $30{\degr}$.

While it is notoriously difficult to estimate the errors in maps reconstructed with ZDI, previous studies applied a bootstrap technique to test the robustness of the magnetic images \citep[e.g.,][]{ZDK21,ZDK22,SCI23}. These works suggest error bars typically equal to 5$\%$ for field strengths, and up to 10{\degr} on the field inclinations. Assuming similar uncertainties, these results indicate that the minor changes in our magnetic maps are real and mostly reflect a dipolar field excursion.
 
For completeness, we summarised in Table~\ref{tab:mag_properties} the main properties of the magnetic field geometries derived in this study. We provide the maximum magnetic field strength at the stellar surface ($B_\mathrm{max}$), and the fractional energy stored in the poloidal ($E_\mathrm{Pol}$) and toroidal ($E_\mathrm{Tor}$) field components. In addition, we report the distribution of poloidal field energy into dipolar ($E_\mathrm{Dip}$), quadrupolar ($E_\mathrm{Quad}$), and octupolar ($E_\mathrm{Oct}$) components. The fraction of poloidal energy on axisymmetric modes and the level of axisymmetry of the dipole field ($E_\mathrm{Dip}$) are also displayed in Table~\ref{tab:mag_properties}, along with the tilted dipole field strength ($B_\mathrm{Dip}$) and obliquity $\beta$.

\begin{table}
	\centering
	\caption{Magnetic field properties of GM~Aur derived with ZDI using data set $\#1$ (second column), data set $\#2$ (third column), and data set $\#3$ (fourth column).}
	\label{tab:mag_properties}
	\begin{tabular}{llll} 
		\hline
		 Magnetic field properties &  Data set $\#1$ & Data set $\#2$ & Data set $\#3$\\
 		\hline  
       $B_\mathrm{max}$  (G) & $880$ & $900$ & $980$\\
       $E_\mathrm{Pol}$  ($E_\mathrm{Tot}$) & $99\%$  & $99\%$ & $98\%$ \\
       $E_\mathrm{Tor}$  ($E_\mathrm{Tot}$) & $1\%$   & $1\%$ & $2\%$ \\
       $E_\mathrm{Dip}$  ($E_\mathrm{Pol}$) & $65\%$  & $65\%$ & $62\%$ \\
       $E_\mathrm{Quad}$ ($E_\mathrm{Pol}$) & $1\%$   & $1\%$ & $1\%$\\
       $E_\mathrm{Oct}$  ($E_\mathrm{Pol}$) & $33\%$   & $33\%$ &  $36\%$ \\
       Pol. axisymmetric       ($E_\mathrm{Pol}$) & $98\%$  & $98\%$ & $96\%$ \\
       $B_\mathrm{Dip}$  (G)  & $740$ & $750$ & $705$\\
       $\beta$      ($\degr$) & $15$ & $10$ & $10$\\
       Rotation phase of tilted dipole & 0.91  & 0.95 & 0.81\\
\noalign{\smallskip}
		\hline
    \end{tabular}
\end{table}

\section{Radial velocity}\label{sec:GPrv}
We estimated RV variations of GM Aur by computing a Gaussian fit to the Stokes $I$ LSD profiles from atomic lines and those from CO bandhead lines (see Table~\ref{tab:results}). Our measurements using LSD profiles from atomic lines show a mean value \vrad $ \equiv \left< RV \right>$  of $14.94$\kmps and a standard deviation of $0.31$\kmps, which is consistent with earlier spectroscopic estimations of $14.95 \pm 0.98$\kmps \citep{MBG20}. In contrast, the RVs obtained from CO lines are shifted by $0.63$\kmps with respect to the atomic lines. They feature a mean value of $15.57$\kmps and a standard deviation of $0.27$\;km\;s$^{-1}$. We observe a statistically significant positive correlation, represented by a Pearson correlation coefficient of 0.87, between the RVs obtained from atomic lines and CO lines.

\begin{table*}
    \caption[]{MCMC results for the joint modelling of the RV data from atomic and CO lines of GM~Aur. Results correspond to the GP models with and without planet. Note that the values of $\eta_2$, $\eta_3$, and $\eta_4$ are shared in the GPs of atomic and CO lines.}
    \label{tab:gpRVjointfitparams}
    \begin{tabular}{lllll}
        \hline
     Parameters & Name & GP model values & GP+Planet model values & Prior \\
        \hline
     GP semi-amplitude of the atomic lines   (km\,s$^{-1}$)  & $\eta_1$ (atomic lines)    & $0.33^{+0.07}_{-0.05}$ & $0.32^{+0.06}_{-0.05}$ & $\mathcal{U}(0,100)$ \\
        \noalign{\smallskip}
     GP semi-amplitude of the CO lines  (km\,s$^{-1}$)  & $\eta_1$ (CO lines)    & $0.24^{+0.06}_{-0.04}$ & $0.23^{+0.05}_{-0.04}$ & $\mathcal{U}(0,100)$ \\
        \noalign{\smallskip}
     Rotation period (d) & $\eta_2$ & $6.00\pm0.01$ & $5.99\pm0.01$  & $\mathcal{N}(6.04,1)$ \\
        \noalign{\smallskip}
     Decay time scale (d)  & $\eta_3$     & $97  $  & $97  $ & FIXED \\
        \noalign{\smallskip}
     Smoothing factor      & $\eta_4$     & $0.33$  & $0.33$ & FIXED \\
        \noalign{\smallskip}
     White noise of the atomic lines (km\,s$^{-1}$)    & $\eta_5$  (atomic lines)   & $0.06\pm0.03$  & $0.00\pm0.05$ & $\mathcal{U}(-100,100)$ \\
        \noalign{\smallskip}
     White noise of the CO lines (km\,s$^{-1}$)    & $\eta_5$ (CO lines)     & $0.09\pm0 .03$  & $0.00\pm0.05$ & $\mathcal{U}(-100,100)$ \\
     \noalign{\smallskip}     
     \noalign{\smallskip}
     \hline
     Orbital semi-amplitude    (km\,s$^{-1}$)  & $K_b$     &     & $0.11\pm0.02$ & $\mathcal{U}(0,1)$ \\
        \noalign{\smallskip}
     Orbital period (d)  & $P_b$    &     & $8.745\pm0.009$ & $\mathcal{N}(8.75,0.1)$  \\
          \noalign{\smallskip}
    Time of inferior conjunction (+2459000) & BJD$_b$  &       & $704.0\pm0.5$ & $\mathcal{N}(703.8,2)$ \\
          \noalign{\smallskip}
   Minimum planet mass ($M_\mathrm{Jup}$) & $M_b\sin{i}$  &       & $1.10 \pm 0.30$ & derived from $K_b$, $P_b$, and $M_\star$\\
    \hline     
     Reduced-$\chi^2$         &   $\chi^2_r$  &   1.07  &  0.61 & \\
     RMS of atomic lines (km\,s$^{-1}$) &   RMS (atomic lines) &   0.091  &  0.076 & \\
     RMS of CO lines  (km\,s$^{-1}$) &   RMS (CO lines)  &   0.113  &  0.079 & \\
    \hline
     Marginal log-likelihood  &   $\log\mathcal{L}_M$  &    21.7   & 38.5 & \\
    log-Bayes factor$\,\,^a$  &   $\log\mathrm{BF} = \Delta\log\mathcal{L}_M$  &    0.0   & 16.8 & \\
     \hline
     \multicolumn{5}{l}{$^a$ The Bayes factor is computed with respect to the reference model without a planet.}\\
    \end{tabular}
\end{table*}

Similar to the procedure described in Sec.~\ref{sec:GPbl}, we jointly model the RV curves from atomic lines and CO lines using GP regression. The GP models from atomic and CO lines share a common periodicity found to be $\eta_2 = 6.00\pm0.01$\;d that is consistent with older literature determinations of 5.8–6.1\;d for the stellar rotation period \citep[][]{PGS10,AGP12,RER22,BSP23}. The RV periodicity is however slightly lower than our previous estimate from $\bl$ measurements (of $6.03^{+0.03}_{-0.04}$\;d), which is similar to what was reported for other CTTSs observed with SPIRou \citep[e.g., CI~Tau and TW~Hya,][]{DFC24,DCL24} and likely results from weak levels of surface differential rotation at the surface of GM~Aur. One notable difference between both GP models is that the GP RV curve from atomic lines has a semi-amplitude of $0.33$\;km\;s$^{-1}$, whereas that from CO lines features a lower semi-amplitude of $0.24$\;km\;s$^{-1}$. The difference can be attributed to magnetic fields inducing dominant profile distortions in atomic lines (but not in the magnetically insensitive CO lines) and brightness features generating only weaker perturbations (in both atomic and CO lines). Note that $\eta_3$ and  $\eta_4$ were fixed to their optimal values obtained in the GP analysis of the $\bl$ data (Sec.~\ref{sec:GPbl}), as the limited coverage of our observations does not effectively constrain these parameters from the noisier RV data. 

\begin{figure*}
	\includegraphics[width=\textwidth]{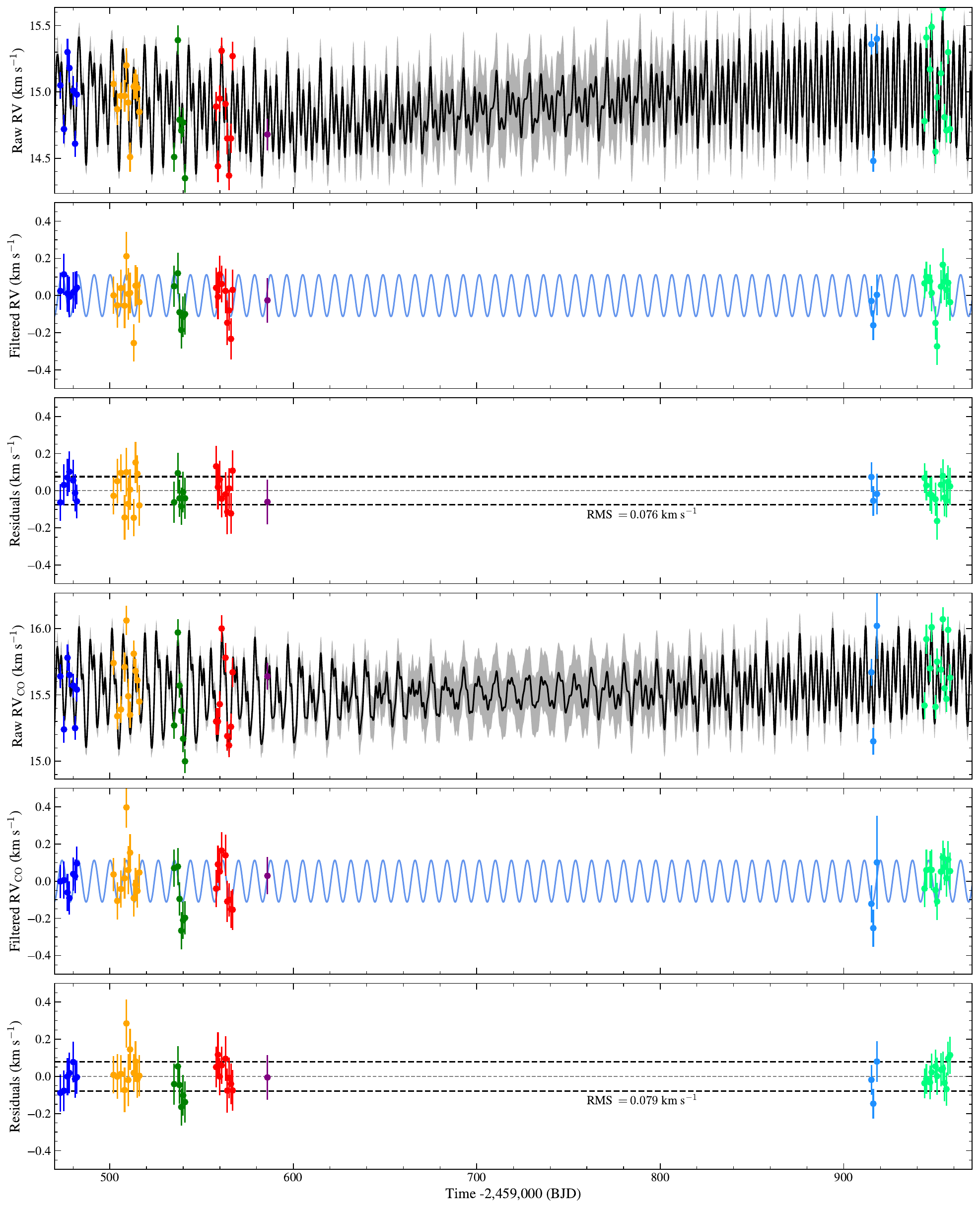}
     \caption{Joint GP regression of the raw RVs from both atomic lines (first 3 rows) and CO lines (last 3 rows) of GM Aur. Rows 1 and 4 illustrate the raw RVs alongside the GP+Planet model prediction. Rows 2 and 5 show to the activity-filtered RV data and the modelled planetary RV signal (blue curve). Rows 3 and 6 give activity- and planet-filtered RV residuals. Symbol colors carry the same meaning as in Fig.~\ref{fig:GPbl}.}
    \label{fig:GPperRVCO}
\end{figure*}

As we detect residual power in the filtered RVs around a period of $8.75$\;d (corresponding to a false alarm probability, $\text{FAP}\sim 0.2\%$), we launched a new MCMC run, including this time an RV signal induced by a putative close-in planet on a circular orbit – i.e., with three more parameters being fitted in the GP model: the planet semi-amplitude, orbital period, and time of inferior conjunction. The results of both fits (with and without planet) are listed in Table~\ref{tab:gpRVjointfitparams}. We find a significant RV signal present in the activity-filtered data at a $5.5\sigma$ level (Fig.~\ref{fig:GPperRVCO}), featuring a semi-amplitude of $0.11\pm0.02$\kmps and an orbital period of $8.745\pm0.009$\;d. The detected RV signal is supported by the log-Bayes factor (or log-marginal likelihood increase) of 16.8 – i.e., well above the minimal detection threshold of 5 suggested by \citet{J83}. Using the GP model with a planet to compute the activity-filtered RVs reveals a periodic signal around $8.745$\;d (see Fig.~\ref{fig:PER_RVCO_Planet}) that is consistent with the residual power seen in the activity only GP model.

\begin{figure}
	\includegraphics[width=\columnwidth]{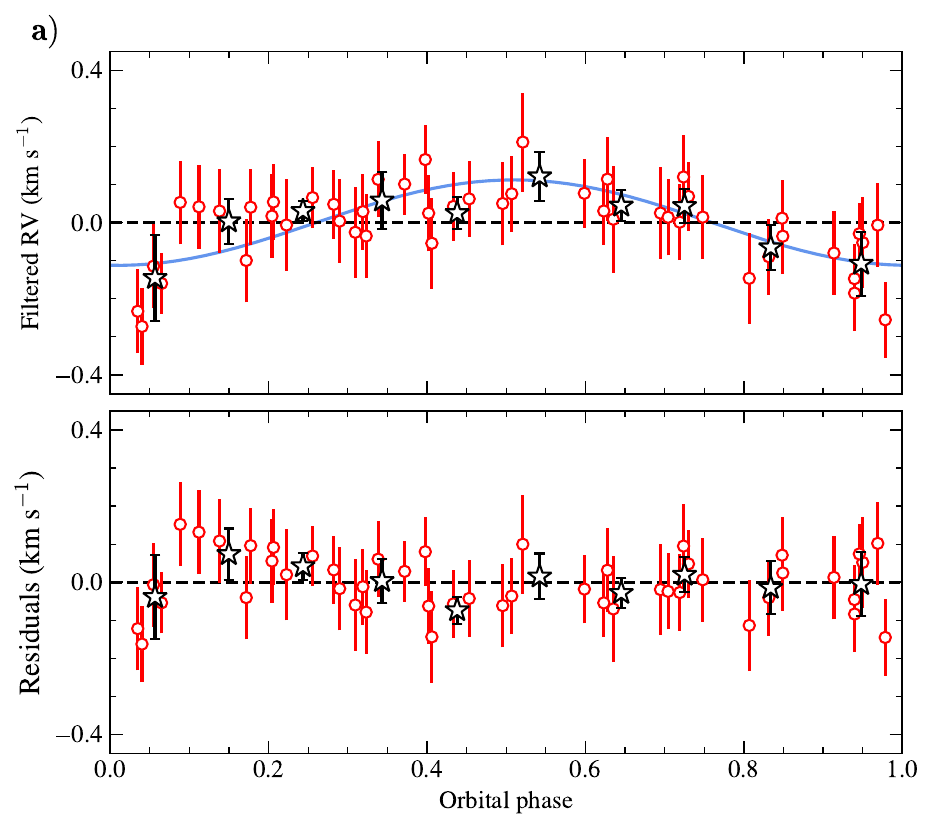}
	\includegraphics[width=\columnwidth]{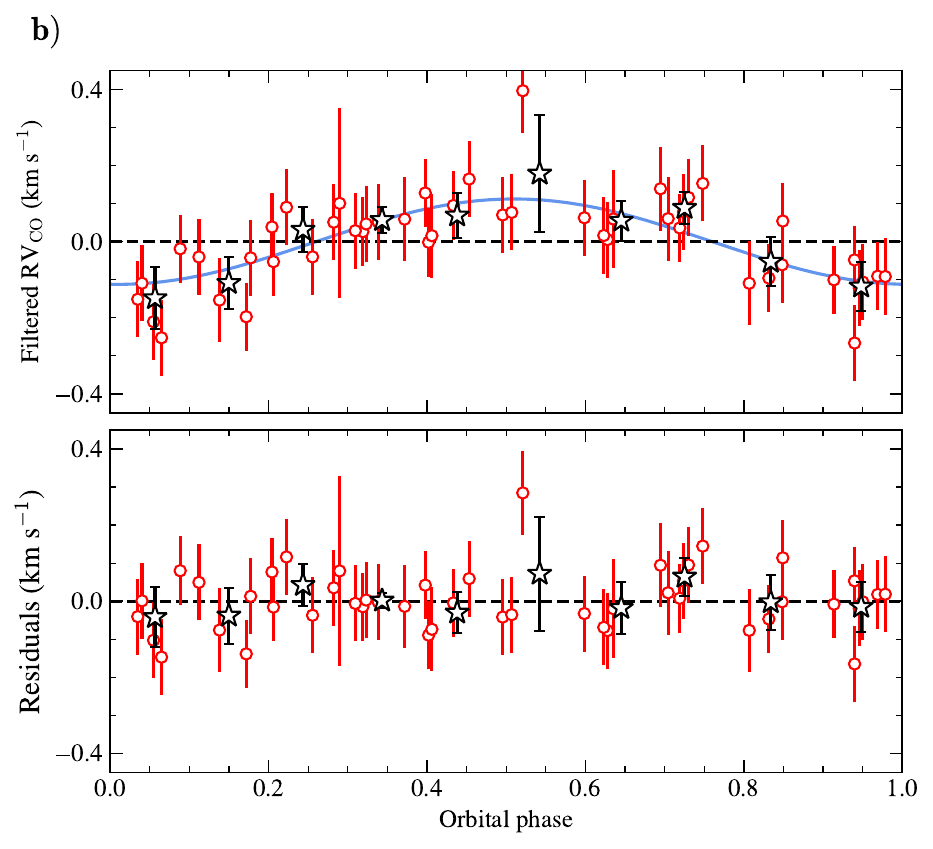}
     \caption{Activity-filtered and residual RV data of GM Aur, respectively for atomic lines ({\bf a}) and for CO lines ({\bf b}), phase-folded on the 8.745\;d orbital period. Red points illustrate the RVs, whereas black stars correspond to averaged values over 0.1 phase bins. The blue sinusoid is the result of the combined MCMC fit to both atomic and CO lines.}
    \label{fig:GPfoldedRVCO}
\end{figure}

If attributed to a candidate planet, the detected signal would imply a minimum planet mass of $M_b\sin{i} = 1.10 \pm0.30\;M_\mathrm{Jup}$, and a planet mass of $M_b = 1.38 \pm0.37\;M_\mathrm{Jup}$ if the planet orbits in the plane of the disc. Fig.~\ref{fig:GPfoldedRVCO} shows the phase-folded activity-filtered RV curve obtained in our modelling for both atomic and CO lines. The posterior distributions obtained in the 8-parameter fit to the joint RV data from atomic lines and CO lines are illustrated in Fig.~\ref{fig:GP_cornerRVjoint}. As evidenced by the stacked periodogram in Fig.~\ref{fig:STACKEDPER_RVCO_Planet}, the RV signal power-detection increases with the number of observations, which is expected for planet-induced signals. We also investigated whether an eccentric orbit yields a better fit than a circular orbit to our RV data, and found an eccentricity consistent with 0 (with an error bar of $\pm0.05$), along with a non significant change in marginal likelikhood (with respect to the circular case), supporting our initial assumption of a circular orbit.

We double-checked that the orbital solutions obtained through an independent fit to the RV data from atomic and CO lines are in good agreement with the solution obtained from the joint modelling of both sets of lines, although with lower Bayesian evidence (of 6.2 and 11.5 for atomic lines and CO lines, respectively). This result tells in particular that the detected RV signal shows up in the whole spectrum and on all spectral lines (as expected from, e.g., a planet RV signal), and not just in one single set of lines or spectral band (if caused by, e.g.,  activity or a non-axisymmetric inner disc structure). Tables~\ref{tab:gpRVatomicfitparams} and \ref{tab:gpRVcofitparams} present the GP results obtained in the individual fit to the RV data from atomic and CO lines, respectively.

\begin{table*}
    \caption[]{Similar to Table~\ref{tab:gpRVjointfitparams}, but for the RV data from atomic lines of GM~Aur.}
    \label{tab:gpRVatomicfitparams}
    \begin{tabular}{lllll}
        \hline
     Parameters & Name & GP model values & GP+Planet model values & Prior \\
        \hline
     GP semi-amplitude    (km\,s$^{-1}$)  & $\eta_1$     & $0.33^{+0.07}_{-0.05}$ & $0.32^{+0.07}_{-0.05}$ & $\mathcal{U}(0,100)$ \\
        \noalign{\smallskip}
     Rotation period (d) & $\eta_2$ & $6.00\pm0.02$ & $5.99\pm0.02$  & $\mathcal{N}(6.04,1)$ \\
     Decay time scale (d)  & $\eta_3$     & $97  $  & $97  $ & FIXED \\
     Smoothing factor      & $\eta_4$     & $0.33$  & $0.33$ & FIXED \\
     White noise    (km\,s$^{-1}$)    & $\eta_5$     & $0.06\pm0.03$  & $0.00\pm0.05$ & $\mathcal{U}(-100,100)$ \\
     \noalign{\smallskip}
     \hline
     Orbital semi-amplitude    (km\,s$^{-1}$)  & $K_b$     &     & $0.09\pm0.03$ & $\mathcal{U}(0,1)$ \\
        \noalign{\smallskip}
     Orbital period (d)  & $P_b$    &     & $8.74^{+0.02}_{-0.03}$ & $\mathcal{N}(8.75,0.1)$  \\
          \noalign{\smallskip}
    Time of inferior conjunction (+2459000) & BJD$_b$  &       & $704\pm1$ & $\mathcal{N}(703.8,2)$ \\
          \noalign{\smallskip}
   Minimum planet mass ($M_\mathrm{Jup}$) & $M_b\sin{i}$  &       & $0.85 \pm 0.30$ & derived from $K_b$, $P_b$, and $M_\star$\\
     \hline     
     Reduced-$\chi^2$           &   $\chi^2_r$  &   0.86  &  0.58 & \\
     Residuals root-mean-square  (km\,s$^{-1}$) &   RMS  &   0.091  &  0.074 & \\
     \hline     
     Marginal log-likelihood  &   $\log\mathcal{L}_M$  &    10.5   & 16.7 & \\
    log-Bayes factor  &   $\log\mathrm{BF} = \Delta\log\mathcal{L}_M$  &    0.0   & 6.2 & \\
     \hline
    \end{tabular}
\end{table*}

\begin{table*}
    \caption[]{Similar to Table~\ref{tab:gpRVjointfitparams}, but for the RV data from CO lines of GM~Aur.}
    \label{tab:gpRVcofitparams}
    \begin{tabular}{lllll}
        \hline
     Parameters & Name & GP model values & GP+Planet model values & Prior \\
        \hline
     GP semi-amplitude    (km\,s$^{-1}$)  & $\eta_1$     & $0.24^{+0.06}_{-0.04}$ & $0.23^{+0.05}_{-0.04}$ & $\mathcal{U}(0,100)$ \\
        \noalign{\smallskip}
     Rotation period (d) & $\eta_2$ & $6.00^{+0.03}_{-0.02}$ & $5.99\pm0.02$  & $\mathcal{N}(6.04,1)$ \\
     Decay time scale (d)  & $\eta_3$     & $97  $  & $97  $ & FIXED \\
     Smoothing factor      & $\eta_4$     & $0.33$  & $0.33$ & FIXED \\
     White noise    (km\,s$^{-1}$)    & $\eta_5$     & $0.09\pm0.03$  & $0.00\pm0.05$ & $\mathcal{U}(-100,100)$ \\
     \noalign{\smallskip}
     \hline
     Orbital semi-amplitude    (km\,s$^{-1}$)  & $K_b$     &     & $0.13\pm0.03$ & $\mathcal{U}(0,1)$ \\
        \noalign{\smallskip}
     Orbital period (d)  & $P_b$    &     & $8.75\pm0.01$ & $\mathcal{N}(8.75,0.1)$  \\
          \noalign{\smallskip}
    Time of inferior conjunction  (+2459000) & BJD$_b$  &       & $704.1^{+0.7}_{-0.8}$ & 
    $\mathcal{N}(703.8,2)$\\
          \noalign{\smallskip}
   Minimum planet mass ($M_\mathrm{Jup}$) & $M_b\sin{i}$  &       & $1.23 \pm 0.30$ & derived from $K_b$, $P_b$, and $M_\star$\\
     \hline     
     Reduced-$\chi^2$           &   $\chi^2_r$  &   1.3  &  0.66 & \\
     Residuals root-mean-square  (km\,s$^{-1}$) &   RMS  &   0.113  &  0.078 & \\
     \hline     
     Marginal log-likelihood  &   $\log\mathcal{L}_M$  &    11.2   & 22.7 & \\
    log-Bayes factor  &   $\log\mathrm{BF} = \Delta\log\mathcal{L}_M$  &    0.0   & 11.5 & \\
     \hline
    \end{tabular}
\end{table*}

Furthermore, it is unlikely that the candidate planet would drive a pulsed accretion mechanism in GM~Aur given its circular orbit \citep{AL96}. In agreement with this, a quick inspection of previously published TESS data does not show significant power at the orbital period.

\section{Summary and discussion} \label{sec:discussion}
We analyzed nIR spectra of the CTTS GM~Aur, collected with the SPIRou spectropolarimeter at CFHT from September 2021 to January 2023. We computed Stokes $I$ and $V$ LSD profiles for the 49 collected spectra. Focusing first on the temporal analysis of the surface longitudinal magnetic field, we detected a quasi-periodic signal with a periodicity of $6.03^{+0.03}_{-0.04}$\,d. This agrees with the range of stellar rotation periods documented in the literature, going from 5.8 to 6.1 days, as determined through spectroscopy and photometry \citep{PGS10,RER22,BSP23}. 

The primary aim of our investigation was to recover the large-scale surface magnetic field of GM~Aur and investigate its magnetospheric accretion regime. To accomplish that, we analyzed the LSD profiles using ZDI, dividing our data into three distinct sets, assuming that surface  brightness (i.e. temperature) inhomogeneities and/or large-scale magnetic fields remained static within each set.

\subsection{The large-scale magnetic field topology}
The reconstructed large-scale magnetic field of GM~Aur revealed a topology predominantly characterized by a dipolar configuration that is slightly tilted towards rotational phases $0.81$-$0.95$. The dipole tilt angles obtained in our study ($10{\degr}$ and $15{\degr}$) agree with the small value reported in November $2011$ by \citet{MBG20}, who inferred a tilt angle $\beta = 13{\degr}^{+16{\degr}}_{-13{\degr}}$ through the analysis of the variability of the optical accretion line HeI\,$587.6$\,nm. The average field strength of this dipole is about $730$\,G, close to those of other CTTSs with similar stellar rotation periods like, e.g., V2129~Oph \citep[of 625\,G,][]{DJG07,DBW11}. Indeed a positive correlation between $B_\mathrm{Dip}$ and \prot has been reported in the literature for CTTS and taken as a natural consequence of star-disc interaction torques \citep{JJG14,VGJ14,AM23}, in which the magnetospheric interaction with the circumstellar disc sets the stellar rotation period. Combining the peak magnetic field at the stellar surface with the magnetic filling factor of about 40 per-cent suggests that the small-scale field may locally reach strengths up to 2.5\,kG, consistent with previous small-scale field measurements from Zeeman broadening of the atomic lines of GM~Aur \citep{JK07,FCR22}.

Our ZDI analysis showed that magnetic effects can alone explain the modulation of the Stokes $I$ and $V$ LSD profiles of atomic lines without the need for brightness inhomogeneities. Indeed, the Stokes $I$  profiles show the most significant distortions in the latest observations when the maximum magnetic field strength is about $100$\,G larger than in the latest epochs (see Fig.~\ref{fig:bestentropyfit}). Brightness inhomogeneities are nonetheless present at the surface of the star (hence the activity jitter in the RV curve of the magnetically insensitive CO lines, see Sec.~\ref{sec:GPrv}), but are found to generate profile distortions in atomic lines smaller than those from magnetic fields. Nevertheless, the companion study conducted by \citet{BSP23} demonstrated the existence of an accretion spot positioned at rotational phase $0$ during the same period of our observations (concomitant to the first two data sets). We argue that this discrepancy most likely arises because we are blind to the accretion spot in the nIR domain, which is defined by typical temperatures of about $8000$\,K \citep{HHC16}. Previous optical studies have resorted to accretion-powered emission lines to gather more information about the hot spot and precisely pinpoint its location at the stellar surface \citep[e.g.,][]{DSB10b,DGA13}. Unfortunately, we are unable to employ a similar procedure with our data as there is no clear accretion proxy that solely probes the accretion spot at the footpoint of accretion funnels in the nIR domain, but rather accretion proxies that carry information from all over the accretion funnel and inner disc regions \citep[e.g.,][]{SBA23}. Contemporaneous nIR and optical spectropolarimetry would be required to draw a picture of the surface brightness distribution of  GM~Aur in future observations. Last but not least, we note that the large-scale field of GM~Aur is tilted towards phase 0.9. This is in rough agreement with the result of \citet{BSP23} reporting that the accretion column (and the chromospheric hot spot at the base of the accretion funnel) is apparently anchored/located at the surface of GM~Aur near phase 0.0.

\begin{figure*}
    \centering
    \includegraphics[width=0.95\textwidth]{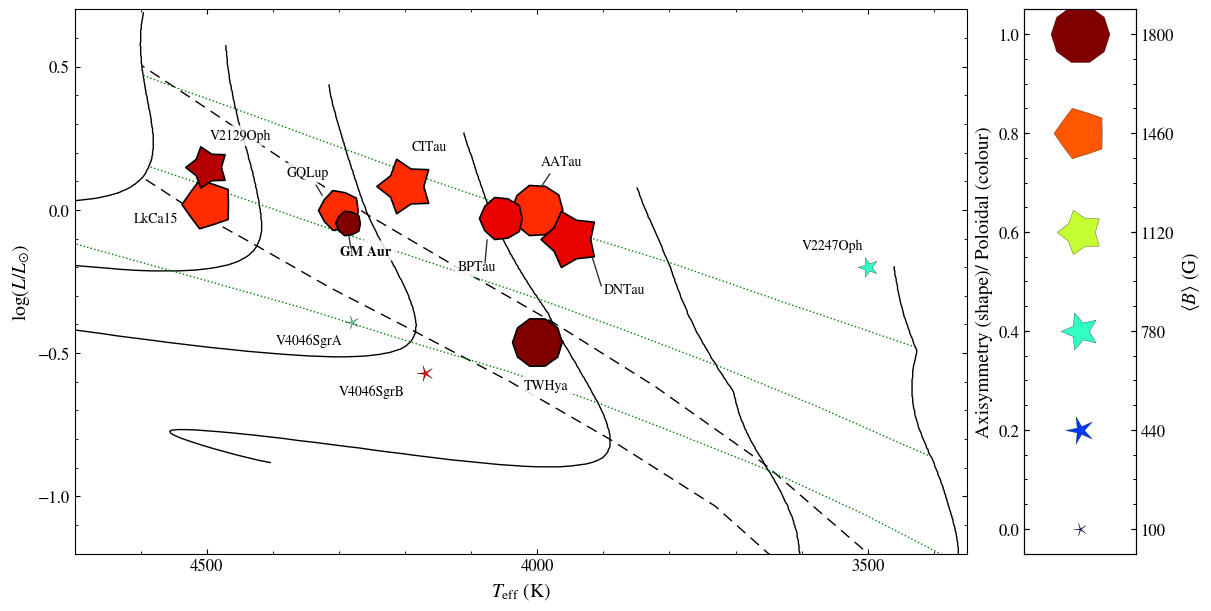}
    \caption{Magnetic Hertzsprung–Russell diagram for CTTS. Symbols depict the properties of the large-scale magnetic field reconstructed for each star; symbol sizes are proportional to the averaged field strength ($\left<B\right>$), colors represent the field configuration (from red to blue for purely poloidal to purely toroidal fields), and shapes illustrate the degree of axisymmetry of the poloidal field component (with higher axisymmetry shown as a more circular symbol). Evolutionary models are similar to Fig.~\ref{fig:hrd}. Mass tracks go from 0.3 to 1.3\,M$_\odot$ in steps of 0.2\,M$_\odot$ (black lines) and isochrones are for ages of $1$, $5$, and $10$\,Myr (dotted green lines).}
    \label{fig:hrdmag}
\end{figure*}

In a broader context, the magnetic topology of GM~Aur supports the work of other studies in this area, linking the magnetic field complexity with the stellar internal structure \citep[e.g.,][]{GDM12}. Figure~\ref{fig:hrdmag} compares the large-scale magnetic field morphology of CTTS at different locations in the Hertzsprung-Russel diagram. This figure shows that the dominant axisymmetric poloidal field of GM~Aur resembles the magnetic configuration of CTTS with either a small radiative core or a fully convective structure (e.g., AA~Tau, BP~Tau, and DN~Tau). GM~Aur's simple large-scale magnetic field morphology translates in a dipole-to-octupole ratio of about 2 at all epochs. This result reinforces that the reconstructed magnetic maps of GM~Aur differ mainly because of the magnetic excursion of the dipolar field component. The minor rearrangement of the field configuration is likely what causes the short-term variability seen in the $\bl$ data. These results hint towards a non-stationary dynamo state, although a clear picture can only be obtained through regular magnetic monitoring of GM~Aur. We emphasize that future analogous works modelling the magnetic field of other CTTSs are of paramount importance to get statistically significant correlations between the magnetic field morphology and the evolutionary stage of stars.

\subsection{GM~Aur's magnetospheric accretion regime}
Previous theoretical investigations have highlighted the significance of the relative positions of the disc corotation radius ($r_\mathrm{cor}$) and the disc truncation radius ($r_\mathrm{m}$) in determining the type of stellar accretion regime \citep[e.g.,][]{BRL16,RBU18}. Based on the polar field strength $B_\mathrm{Dip}$ derived  from the magnetic topologies obtained in this study and considering the accretion rate of $(0.5 \pm 0.4)\times 10^{-8}M_{\odot}\mathrm{yr}^{-1}$ from the H$\beta$ line flux \citep{BSP23}, an estimation of the magnetospheric truncation radius can be made using the analytical solution proposed by \citet{BZF08}:
\begin{equation} \label{eq:trunc}
\frac{r_\mathrm{m}}{R_\star} = 2m_s^{2/7}\left(\frac{B_\mathrm{Dip}}{280\,G}\right)^{4/7}\left(\frac{\dot{M}_\mathrm{acc}}{10^{-8}M_{\odot}\mathrm{yr}^{-1}}\right)^{-2/7}\left(\frac{M_\star}{0.8 M_{\odot}}\right)^{-1/7}\left(\frac{R_\star}{2R_{\odot}}\right)^{5/7}
\end{equation}
where $m_s\approx 1$ is the sonic Mach number, $\dot{M}_\mathrm{acc}$ is the mass accretion rate, and $B_\mathrm{Dip}$ is the dipole field strength at the pole (roughly taken as the mean value from the dipolar field strengths reported in Table~\ref{tab:mag_properties}). Utilizing this equation, the magnetosphere is predicted to truncate the circumstellar disc of GM~Aur at an approximate value of $r_\mathrm{m} = 4.1 \pm 1.0$\,R$_\star$ ($0.039 \pm 0.009$\;au). Furthermore, assuming Keplerian rotation for the inner disc of GM~Aur, as suggested by the CO emission from the disc \citep[e.g.,][]{DGP98,SDG00,HHW13}, the radius at which the angular velocity of the disc matches that of the star can be computed as
\begin{equation}
    r_\mathrm{cor} = \left(\frac{GM_{\star}}{\Omega_\star^2}\right)^{1/3}, \,\,\, \text{where}  \,\,\,\,\,\, \Omega_\star = \frac{\vsin}{R_\star\sin i}.
\end{equation}
This relationship leads to $r_\mathrm{cor} =  6.8 \pm 0.2$\,R$_\star$ ($0.064 \pm 0.002$\;au).

It is possible to draw a picture of the magnetospheric accretion of GM~Aur based on the ratio of $r_\mathrm{m}/r_\mathrm{cor}$ equal to $0.60\pm0.15$. This ratio is at the limit of the stable accretion regime \citep{BRL16,PZB20}, and is apparently enough to generate stable accretion in the system, but not to enforce the star to spin down to the longer rotation periods (8-10\;d) of prototypical CTTSs such as AA Tau or CI Tau \citep{DSB10b,DBA20}. This conclusion is supported by the long-lived accretion pattern reported by \citet{BSP23}, who analyzed photometric and spectroscopic data covering 30 rotational cycles. 

Given this low value of $r_\mathrm{m}/r_\mathrm{cor}$, we speculate that GM~Aur may switch at times to an unstable accretion regime, perhaps explaining the quasi-periodic nature of its light curve and the existence of accretion bursts reported in the literature \citep{RE19,RER22,WEK24b}. Past studies have identified that it is not uncommon for CTTSs to change their accretion behaviour, as seen in NGC 2264 \citep{MAG15,SAB16}. These transitions in the accretion regime could be associated with the onset or efficiency of the Rayleigh-Taylor instability that occurs at the interface between the accretion disc and the magnetosphere, which depends on the strength and orientation of the magnetic field \citep{KR13}. Nevertheless, we cannot exclude the possibility that a variable inner disc density contributes to (or even dictates) the changes in mass accretion rate and the quasi-periodic nature of GM~Aur's light curve \citep{EMH19}. These results further justify monitoring the magnetic field of GM~Aur over the next years.

\subsection{Detection of a RV signal}
We obtained radial velocity measurements from the Stokes $I$ LSD profiles computed from either a photospheric atomic line mask or a CO bandhead line mask. Fitting the time-series data with a Gaussian Process revealed a residual signal in the activity-filtered RVs from atomic and CO lines, detected at a $5.5\sigma$ level. The signal has a semi-amplitude of $0.11\pm0.02$\kmps and a periodicity of $8.745\pm0.009$\;d, 
and potentially reflects the presence of a candidate planet on a circular orbit. The competing GP models (with or without planet) showed significant Bayesian evidence for the detected RV signal, with an increase in the marginal log-likelihood of $\Delta \log\mathcal{L}_M = 16.8$. 

Moreover, we find that the significance of the reported RV signal increases with the number of RV data points considered (see Fig.~\ref{fig:STACKEDPER_RVCO_Planet}), consistent with what would be expected from a candidate planetary companion orbiting GM~Aur \citep{MCC17}. We also confirmed that the orbital parameters obtained in the GP fit of the RV data from atomic and CO lines fitted independently agree within $1\sigma$ with the solution obtained from both sets of lines. However, these models have lower Bayesian evidence for the planet detection (of 6.2 and 11.5) due to the smaller amount of information used in the GP regression.

In conclusion, our research has provided compelling evidence of a candidate newborn giant planet orbiting  GM Aur. If attributed to a candidate planet, the detected RV signal would imply a minimum planet mass of $1.10 \pm 0.30\;M_\mathrm{Jup}$, and a planet mass of $1.38 \pm 0.37\;M_\mathrm{Jup}$ if the planet orbits in the plane of the disc. It would imply that this candidate planet orbits at a distance of $0.082 \pm 0.002$\;au ($8.74\pm0.32\;R_\star$), located within the inner accretion disc, slightly beyond the corotation radius. This makes GM Aur b one of the very few close-in candidate planets currently identified around CTTSs \citep[e.g.,][]{MSB24,DFC24}. Finally, it is unlikely that a single planet can carve the large dust cavity observed in GM Aur ($\sim 40$\,au). The odds are that multiple planets exist within the dust gap of GM~Aur, as observed for other stars with large dust cavities such as the CTTS PDS~70 \citep{HBdB19}. Further observations are needed to firmly confirm the planetary nature of the RV signal uncovered in this study and to characterize the planet's atmospheric properties, providing us with an ideal opportunity to derive observational constraints for theoretical models of the formation and evolution of close-in giant planets.

\section*{Acknowledgements}
We thank the anonymous referee for helping improve and clarify the paper. BZ acknowledges funding from the CAPES-PrInt program (\#88887.683070/2022-00 and \#88887.802913/2023-00). BZ and SHPA acknowledge financial support from CNPq, CAPES and FAPEMIG (APQ-01033-22). JFD, BZ, and CM acknowledge funding from the European Research Council (ERC) under the H2020 research \& innovation programme (grant agreement $\#740651$ NewWorlds). JB acknowledges funding from ERC (grant agreement $\#$742095 SPIDI). AC  acknowledge funding from the French ANR under contract number ANR\-18\-CE31\-0019 (SPlaSH). This work is supported by the French National Research Agency in the framework of the Investissements d'Avenir program (ANR-15-IDEX-02), through the funding of the ``Origin of Life" project of the Grenoble-Alpes University.

This paper is based on observations obtained at CFHT which is operated by the National Research Council of Canada, the Institut National des Sciences de l'Univers of the Centre National de la Recherche Scientique of France, and the University of Hawaii. The observations at the CFHT were performed with care and respect from the summit of Maunakea which is a significant cultural and historic site.

\section*{Data Availability}
This paper includes data collected by the SPIRou spectropolarimeter taken as part of the CFHT Large Programs SLS and SPICE (program IDs:  21BP42 $\&$ 22BP45). The SLS data is already publicly available from the Canadian Astronomy Data Center whereas the SPICE data will become available by the second semester of 2025.



\bibliographystyle{mnras}
\bibliography{example} 



\appendix
\section{Log of observations} 
Table~\ref{tab:results} summarizes the log of observations collected by the SPIRou spectropolarimeter from 2021 September to 2023 January, and taken as part of CFHT large programs SLS and SPICE.

\begin{table*}
	\centering
	\caption{Log of the SPIRou observations of GM~Aur collected from 2021 September to 2023 January. Columns 1 and 2, respectively, give the UT data and the barycentric Julian date derived from the mean observation times of the four sub-exposures used to derive a polarimetric sequence. Column 3 provides the rotation cycle $E$ obtained using the ephemeris given by Eq.~\ref{eq:ephemeris} and column 4 the SNR of the polarization sequences per 2.28\kmps bin. Average noise levels of Stokes $V$ LSD profiles with respect to the unpolarized continuum level $I_c$ are shown in column 5. Columns 6 and 7 give the EW \citep[with typical uncertainties of 0.01\kmps, computed as defined in][]{N18} and full width at half maximum (FWHM) of veiled Stokes $I$ LSD profiles, respectively. Column 8 displays the RVs obtained from the centroid of a Gaussian fit to the Stokes $I$ LSD profiles and column 9 the error associated with the determination of the centroid position. Similarly, columns 10 and 11 provide RVs and error bars obtained from the LSD profiles built with the CO line mask. Longitudinal magnetic field measurements and standard deviations (see Eq.~\ref{eq:blong}) are shown in columns 12 and 13, respectively.}
	\label{tab:results} 
	\begin{tabular}{lllllllllllll} 
		\hline
		 Date & Julian Date & E  & SNR & $\sigma_\text{LSD}$ &  EW & FWHM & RV & $\sigma_\mathrm{RV}$  & RV$_\mathrm{CO}$ & $\sigma_\mathrm{RV_{CO}}$  & $\bl$ & $\sigma_{B_\ell}$  \\
\noalign{\smallskip}
		   &  (+2,459,000 d) & & & ($10^{-4}$) &(\kmps) & (\kmps)  & (\kmps) & (\kmps) & (\kmps) & (\kmps) & (G)  &  (G) \\
\hline
	 ($1$)  & ($2$) & ($3$) & ($4$) & ($5$) & ($6$) & ($7$)  & ($8$) & ($9$) & ($10$) &  ($11$) & ($12$) &  ($13$)\\
\noalign{\smallskip}
  \hline
Sept 15, 2021   & $473.06044 $ &  $   2.030 $ & $ 197 $ & $1.78$ & $1.06 $ & $24.67 $ & $ 15.05$ & $0.10$ & $ 15.64$ & $0.11$  & $    -158 $ & $    13$ \\
Sept 17, 2021   & $475.03513 $ &  $   2.357 $ & $ 201 $ & $1.69$ & $1.07 $ & $24.41 $ & $ 14.72$ & $0.11$ & $ 15.24$ & $0.13$  & $     -35 $ & $    13$ \\
Sept 19, 2021   & $476.96335 $ &  $   2.676 $ & $ 231 $ & $1.42$ & $1.01 $ & $24.94 $ & $ 15.30$ & $0.10$ & $ 15.78$ & $0.12$  & $    -108 $ & $    11$ \\
Sept 20, 2021   & $478.01955 $ &  $   2.851 $ & $ 223 $ & $1.43$ & $0.94 $ & $24.63 $ & $ 15.18$ & $0.11$ & $ 15.65$ & $0.12$  & $    -147 $ & $    12$ \\
Sept 22, 2021   & $480.07583 $ &  $   3.191 $ & $ 238 $ & $1.37$ & $1.02 $ & $24.90 $ & $ 15.01$ & $0.11$ & $ 15.57$ & $0.12$  & $     -95 $ & $    11$ \\
Sept 23, 2021   & $481.07949 $ &  $   3.358 $ & $ 225 $ & $1.64$ & $1.02 $ & $24.87 $ & $ 14.61$ & $0.10$ & $ 15.25$ & $0.12$  & $     -30 $ & $    13$ \\
Sept 24, 2021   & $482.07945 $ &  $   3.523 $ & $ 235 $ & $1.36$ & $1.05 $ & $24.66 $ & $ 14.98$ & $0.09$ & $ 15.54$ & $0.12$  & $     -69 $ & $    10$ \\
Oct 14, 2021   & $502.06869 $ &  $   6.833 $ & $ 224 $ & $1.55$ & $0.99 $ & $24.58 $ & $ 15.06$ & $0.10$ & $ 15.74$ & $0.12$  & $    -147 $ & $    12$ \\
Oct 16, 2021   & $504.08913 $ &  $   7.167 $ & $ 215 $ & $1.81$ & $1.00 $ & $24.85 $ & $ 14.87$ & $0.12$ & $ 15.34$ & $0.13$  & $    -111 $ & $    14$ \\
Oct 18, 2021   & $506.07548 $ &  $   7.496 $ & $ 225 $ & $1.50$ & $1.01 $ & $24.44 $ & $ 14.97$ & $0.10$ & $ 15.39$ & $0.12$  & $     -85 $ & $    12$ \\
Oct 20, 2021   & $508.07626 $ &  $   7.827 $ & $ 224 $ & $1.53$ & $0.91 $ & $24.43 $ & $ 14.97$ & $0.12$ & $ 15.71$ & $0.14$  & $    -146 $ & $    14$ \\
Oct 21, 2021   & $509.08032 $ &  $   7.993 $ & $ 223 $ & $1.57$ & $0.88 $ & $24.33 $ & $ 15.20$ & $0.13$ & $ 16.06$ & $0.14$  & $    -134 $ & $    14$ \\
Oct 22, 2021   & $510.07930 $ &  $   8.159 $ & $ 170 $ & $1.97$ & $0.84 $ & $25.18 $ & $ 14.92$ & $0.14$ & $ 15.49$ & $0.16$  & $     -57 $ & $    18$ \\
Oct 23, 2021   & $511.06749 $ &  $   8.322 $ & $ 232 $ & $1.47$ & $1.02 $ & $24.82 $ & $ 14.51$ & $0.11$ & $ 15.35$ & $0.12$  & $     -62 $ & $    12$ \\
Oct 25, 2021   & $513.08220 $ &  $   8.656 $ & $ 182 $ & $1.66$ & $1.00 $ & $25.20 $ & $ 15.04$ & $0.10$ & $ 15.81$ & $0.12$  & $    -119 $ & $    13$ \\
Oct 26, 2021   & $514.04225 $ &  $   8.815 $ & $ 206 $ & $1.64$ & $0.96 $ & $24.78 $ & $ 15.06$ & $0.11$ & $ 15.68$ & $0.12$  & $    -125 $ & $    14$ \\
Oct 27, 2021   & $515.07220 $ &  $   8.985 $ & $ 226 $ & $1.50$ & $1.01 $ & $24.39 $ & $ 15.03$ & $0.10$ & $ 15.61$ & $0.12$  & $    -135 $ & $    12$ \\
Oct 28, 2021   & $516.09521 $ &  $   9.155 $ & $ 192 $ & $1.66$ & $1.02 $ & $25.05 $ & $ 14.85$ & $0.11$ & $ 15.45$ & $0.13$  & $     -99 $ & $    13$ \\
\noalign{\smallskip}
\noalign{\smallskip}
Nov 16, 2021   & $535.09324 $ &  $  12.300 $ & $ 182 $ & $1.79$ & $1.01 $ & $24.52 $ & $ 14.51$ & $0.11$ & $ 15.27$ & $0.13$  & $    -106 $ & $    14$ \\
Nov 18, 2021   & $537.09010 $ &  $  12.631 $ & $ 170 $ & $1.89$ & $1.03 $ & $25.09 $ & $ 15.39$ & $0.11$ & $ 15.97$ & $0.13$  & $     -88 $ & $    15$ \\
Nov 19, 2021   & $538.03226 $ &  $  12.787 $ & $ 219 $ & $1.55$ & $1.03 $ & $24.61 $ & $ 14.79$ & $0.10$ & $ 15.57$ & $0.12$  & $    -105 $ & $    12$ \\
Nov 20, 2021   & $538.97889 $ &  $  12.944 $ & $ 182 $ & $1.77$ & $0.99 $ & $24.67 $ & $ 14.71$ & $0.10$ & $ 15.38$ & $0.13$  & $    -142 $ & $    14$ \\
Nov 21, 2021   & $539.98340 $ &  $  13.110 $ & $ 232 $ & $1.50$ & $1.01 $ & $25.38 $ & $ 14.77$ & $0.11$ & $ 15.17$ & $0.13$  & $    -123 $ & $    12$ \\
Nov 22, 2021   & $541.00639 $ &  $  13.279 $ & $ 221 $ & $1.43$ & $0.96 $ & $24.66 $ & $ 14.35$ & $0.11$ & $ 15.00$ & $0.12$  & $    -121 $ & $    12$ \\
Dec 09, 2021   & $557.97447 $ &  $  16.088 $ & $ 214 $ & $1.51$ & $0.91 $ & $24.98 $ & $ 14.89$ & $0.11$ & $ 15.30$ & $0.13$  & $     -91 $ & $    14$ \\
Dec 10, 2021   & $558.94112 $ &  $  16.249 $ & $ 196 $ & $1.53$ & $0.92 $ & $24.67 $ & $ 14.44$ & $0.12$ & $ 15.30$ & $0.13$  & $    -123 $ & $    14$ \\
Dec 11, 2021   & $559.95316 $ &  $  16.416 $ & $ 166 $ & $1.86$ & $1.00 $ & $24.34 $ & $ 14.95$ & $0.10$ & $ 15.43$ & $0.13$  & $    -111 $ & $    15$ \\
Dec 12, 2021   & $560.95771 $ &  $  16.582 $ & $ 174 $ & $1.89$ & $1.00 $ & $25.31 $ & $ 15.31$ & $0.10$ & $ 16.00$ & $0.12$  & $    -108 $ & $    15$ \\
Dec 14, 2021   & $563.07065 $ &  $  16.932 $ & $ 200 $ & $1.70$ & $0.91 $ & $24.05 $ & $ 14.91$ & $0.12$ & $ 15.78$ & $0.14$  & $    -151 $ & $    15$ \\
Dec 15, 2021   & $564.05049 $ &  $  17.094 $ & $ 185 $ & $1.71$ & $0.91 $ & $25.14 $ & $ 14.65$ & $0.12$ & $ 15.19$ & $0.14$  & $     -79 $ & $    15$ \\
Dec 16, 2021   & $564.98592 $ &  $  17.249 $ & $ 172 $ & $1.70$ & $0.94 $ & $24.30 $ & $ 14.37$ & $0.11$ & $ 15.12$ & $0.12$  & $    -117 $ & $    14$ \\
Dec 17, 2021   & $566.04362 $ &  $  17.424 $ & $ 182 $ & $1.99$ & $1.00 $ & $24.22 $ & $ 14.65$ & $0.11$ & $ 15.26$ & $0.13$  & $     -85 $ & $    16$ \\
Dec 18, 2021   & $566.94675 $ &  $  17.574 $ & $ 190 $ & $1.79$ & $1.02 $ & $25.17 $ & $ 15.27$ & $0.11$ & $ 15.67$ & $0.14$  & $     -96 $ & $    14$ \\
Jan 06, 2022   & $585.93606 $ &  $  20.718 $ & $ 228 $ & $1.47$ & $0.94 $ & $24.22 $ & $ 14.68$ & $0.12$ & $ 15.64$ & $0.12$  & $    -107 $ & $    12$ \\
\noalign{\smallskip}
\noalign{\smallskip}
Dec 01, 2022   & $915.06659 $ &  $  75.210 $ & $ 224 $ & $1.37$ & $1.03 $ & $23.86 $ & $ 15.36$ & $0.08$ & $ 15.67$ & $0.12$  & $     -56 $ & $    10$ \\
Dec 02, 2022   & $916.10661 $ &  $  75.382 $ & $ 209 $ & $1.45$ & $1.03 $ & $24.21 $ & $ 14.48$ & $0.08$ & $ 15.15$ & $0.13$  & $    -118 $ & $    11$ \\
Dec 04, 2022   & $918.07287 $ &  $  75.707 $ & $ 206 $ & $1.71$ & $1.07 $ & $25.90 $ & $ 15.40$ & $0.11$ & $ 16.02$ & $0.32$  & $     -61 $ & $    13$ \\
Dec 30, 2022   & $944.00911 $ &  $  80.002 $ & $ 214 $ & $1.48$ & $1.07 $ & $24.96 $ & $ 14.78$ & $0.08$ & $ 15.42$ & $0.13$  & $    -127 $ & $    11$ \\
Dec 31, 2022   & $945.02590 $ &  $  80.170 $ & $ 207 $ & $1.50$ & $1.08 $ & $25.45 $ & $ 15.41$ & $0.08$ & $ 15.92$ & $0.14$  & $     -56 $ & $    11$ \\
Jan 02, 2023   & $947.01043 $ &  $  80.498 $ & $ 204 $ & $1.45$ & $1.06 $ & $24.23 $ & $ 15.17$ & $0.09$ & $ 15.70$ & $0.13$  & $    -134 $ & $    11$ \\
Jan 03, 2023   & $947.93475 $ &  $  80.651 $ & $ 203 $ & $1.32$ & $1.00 $ & $24.45 $ & $ 15.49$ & $0.10$ & $ 16.01$ & $0.14$  & $     -52 $ & $    10$ \\
Jan 05, 2023   & $949.99284 $ &  $  80.992 $ & $ 206 $ & $1.42$ & $1.06 $ & $24.60 $ & $ 14.55$ & $0.09$ & $ 15.41$ & $0.12$  & $    -127 $ & $    11$ \\
Jan 06, 2023   & $950.87064 $ &  $  81.138 $ & $ 166 $ & $1.93$ & $1.07 $ & $24.48 $ & $ 14.96$ & $0.10$ & $ 15.75$ & $0.13$  & $     -89 $ & $    14$ \\
Jan 08, 2023   & $952.98862 $ &  $  81.488 $ & $ 173 $ & $1.74$ & $1.08 $ & $24.44 $ & $ 15.14$ & $0.09$ & $ 15.69$ & $0.13$  & $    -126 $ & $    13$ \\
Jan 09, 2023   & $953.99865 $ &  $  81.655 $ & $ 189 $ & $1.41$ & $1.07 $ & $25.03 $ & $ 15.63$ & $0.09$ & $ 16.07$ & $0.12$  & $     -52 $ & $    11$ \\
Jan 10, 2023   & $954.95338 $ &  $  81.813 $ & $ 203 $ & $1.41$ & $1.07 $ & $25.25 $ & $ 14.81$ & $0.10$ & $ 15.55$ & $0.14$  & $     -82 $ & $    10$ \\
Jan 11, 2023   & $955.97002 $ &  $  81.982 $ & $ 209 $ & $1.37$ & $1.11 $ & $24.83 $ & $ 14.71$ & $0.09$ & $ 15.47$ & $0.12$  & $    -118 $ & $    10$ \\
Jan 12, 2023   & $956.90486 $ &  $  82.137 $ & $ 201 $ & $1.42$ & $1.09 $ & $24.96 $ & $ 15.30$ & $0.09$ & $ 15.99$ & $0.13$  & $     -68 $ & $    11$ \\
Jan 13, 2023   & $957.94626 $ &  $  82.309 $ & $ 213 $ & $1.33$ & $1.06 $ & $24.70 $ & $ 14.72$ & $0.10$ & $ 15.63$ & $0.13$  & $     -87 $ & $    10$ \\
\noalign{\smallskip}
		\hline
    \end{tabular}
\end{table*}

\section{Unveiled Stokes LSD profiles} \label{appendix:unveil}
We compute unveiled unpolarized ($I$) and polarized ($V$) Stokes LSD profiles (from atomic lines) through the relations:
\begin{equation}
     I = \frac{\mathcal{W}}{EW} I_\text{veil}
\end{equation}
and
\begin{equation}
    V = \frac{\mathcal{W}}{EW} V_\text{veil}.
\end{equation}
Here, veiled unpolarized and polarized LSD profiles are denoted by $I_\mathrm{veil}$ and $V_\mathrm{veil}$, respectively. $EW$ is the equivalent width of the pseudo-line profile $I_\mathrm{veil}$, and $\mathcal{W}$ is a reference equivalent width arbitrarily set to $1.11$\kmps in this paper. Similarly, one can get the SNR of unveiled profiles using the following equations
\begin{equation} \label{eq:unvsnri}
 \text{SNR}_I = \frac{\mathcal{W}}{EW} \text{SNR}_{I\text{veil}},
 \end{equation}
and
\begin{equation} \label{eq:unvsnrv}
     \text{SNR}_V = \frac{\mathcal{W}}{EW} \text{SNR}_{V\text{veil}}.
\end{equation}

\begin{figure*}
	\includegraphics[width=\textwidth]{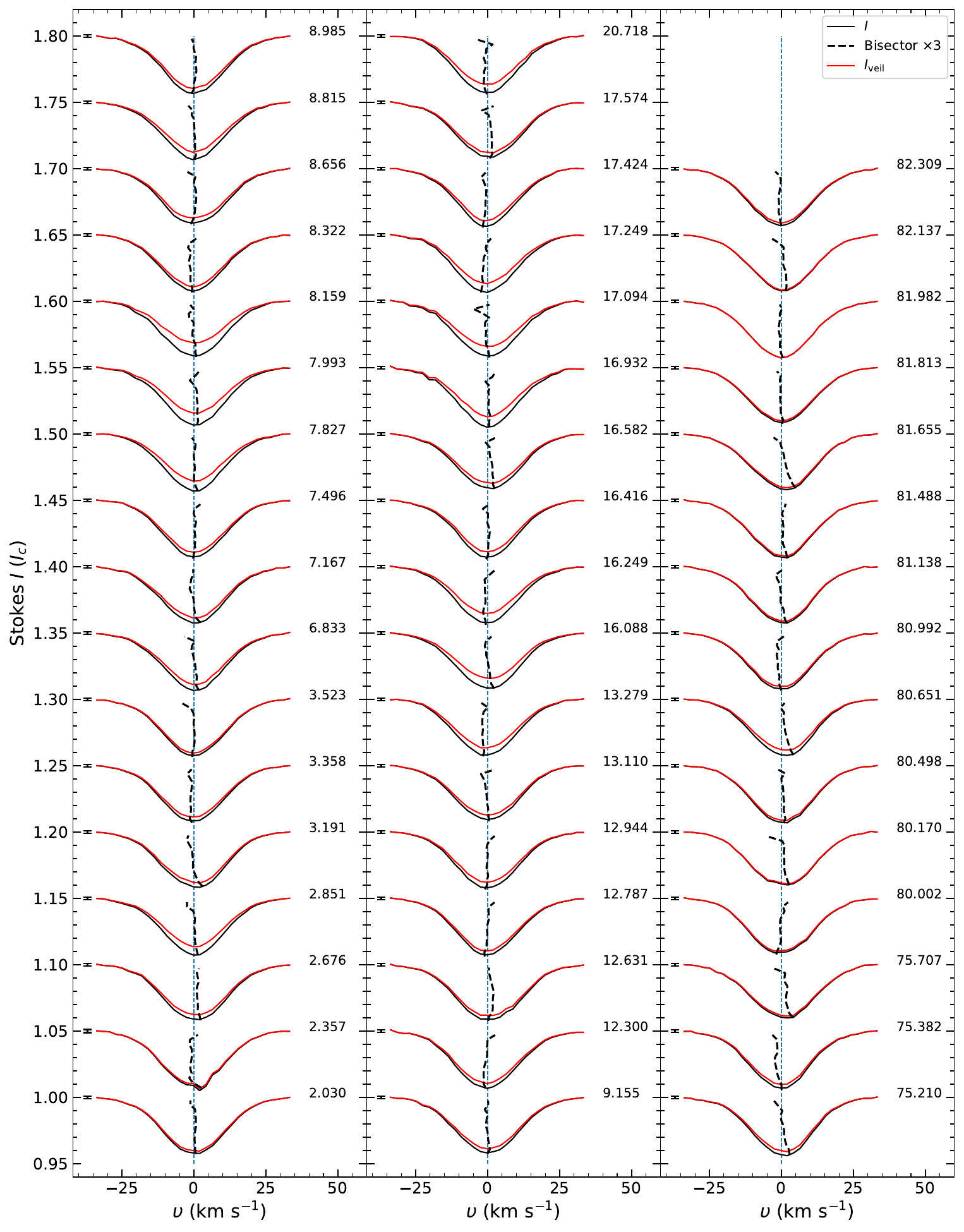}
    \caption{Unveiled (black lines, $I$) and veiled (red lines, $I_\text{veil}$) unpolarized LSD profiles of GM~Aur. Line asymmetries are also evidenced when comparing the reference zero velocity (vertical dashed blue line) to the unveiled line bisectors \citep{G82}. Rotational cycles and 1-$\sigma$ error bars are shown next to each profile.}
    \label{fig:stokesiunveiled}
\end{figure*}

\begin{figure*}
	\includegraphics[width=\textwidth]{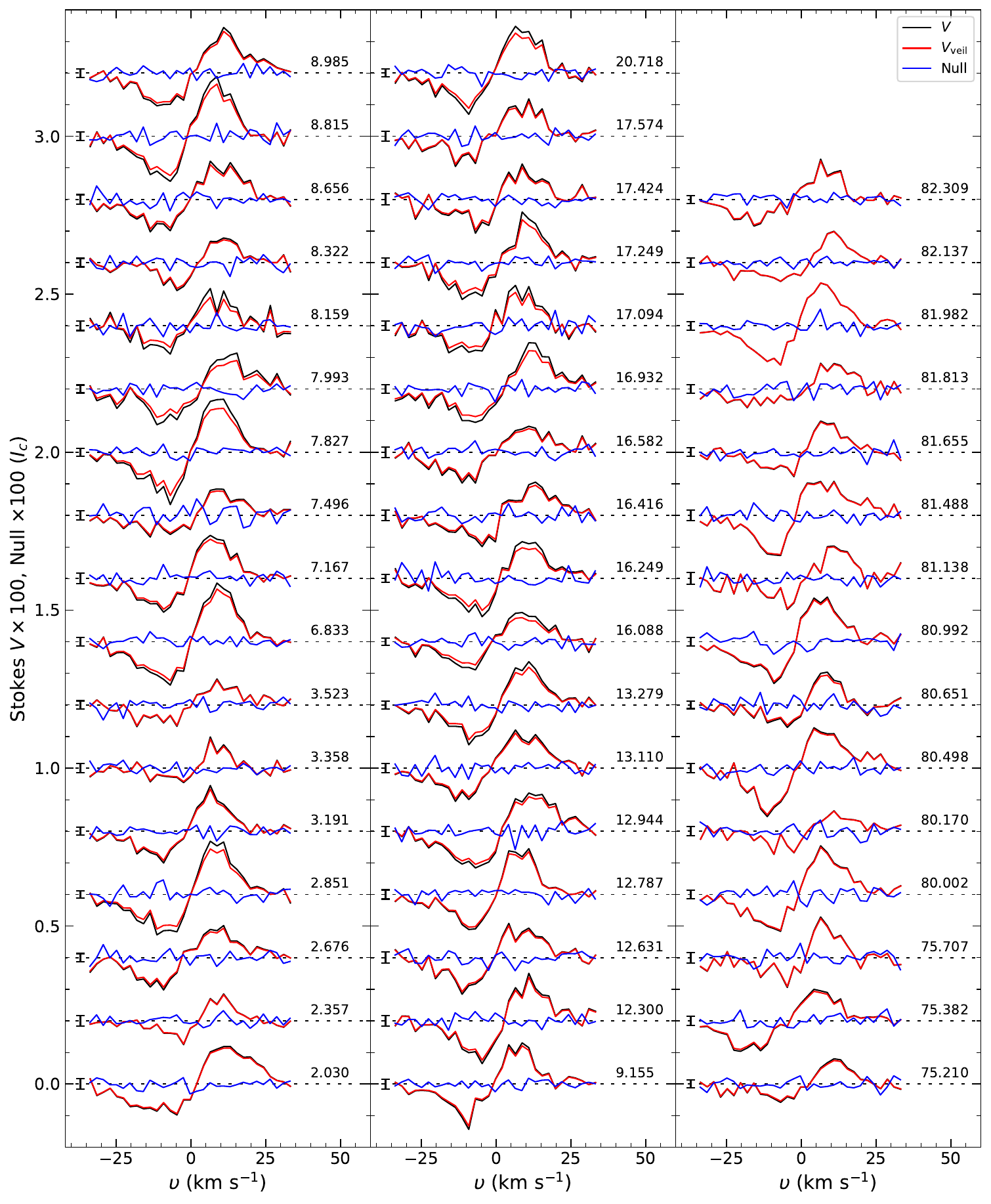}
    \caption{Unveiled (black lines, $V$) and veiled (red lines, $V_\text{veil}$) polarized LSD profiles of GM~Aur. Control null polarisation profiles are shown as blue continuous lines to evidence the lack of spurious polarisation signatures in the Stokes $V$ LSD profiles. Rotational cycles and 1-$\sigma$ error bars (scaled following Eq.~\ref{eq:unvsnrv}) are shown next to each profile.}
    \label{fig:stokesvunveiled}
\end{figure*}

Figures~\ref{fig:stokesiunveiled} and \ref{fig:stokesvunveiled} illustrate, respectively, time-series unpolarized and polarized LSD signatures of GM~Aur before (red lines) and after (black lines) applying the unveiling procedure described above. Comparing both Stokes profiles, it is evident that only minor corrections are needed to mitigate veiling variability at the observation time window. This qualitative finding is consistent with weak veiling variability reported for GM~Aur in the literature \citep{MBG20,LSM21,BSP23}. Line bisectors of the unveiled Stokes $I$ LSD profiles are also displayed in Fig.~\ref{fig:stokesiunveiled} (dashed black line), following the definition of \citet{G82}.  

Finally, Fig.~\ref{fig:stokesvunveiled} also shows null polarisation profiles (blue lines). The noise level null signal provides evidence of the optimal extraction of polarized signatures considered in the paper \citep[see][for a discussion about spurious polarisation signals in cool stars]{FPB16}.

\section{Additional information on the radial velocity analysis} \label{sec:corner_planet}
In Sec.~\ref{sec:GPrv}, we first considered the activity modelling of the raw RV data from atomic and CO lines. Once the stellar activity was filtered out, we could detect a clear periodic signature in the RVs from CO lines at $8.75\pm0.02$\;d. Guided by this periodic signal, we explored a second GP model including the RV wobble induced by a planet in a circular orbit. 

\subsection{Joint fit of the RV data from atomic and CO lines} \label{appendix:jointfit}
\begin{figure*}
	\includegraphics[width=\textwidth]{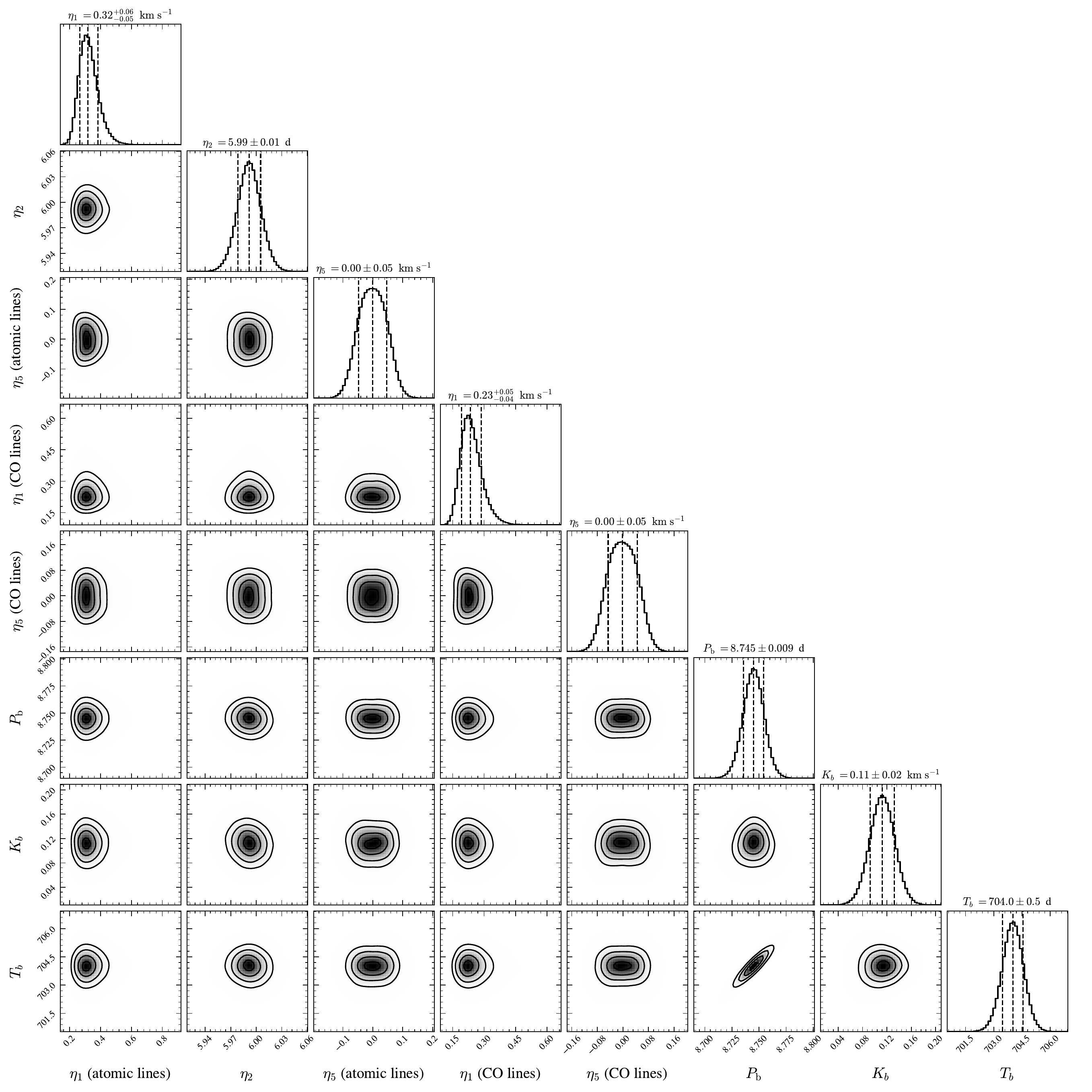}
    \caption{Similar to Fig.~\ref{fig:GP_corner}, but for GP+Planet model of the joint RV data from atomic and CO lines. Semi-amplitude and white noise values are given in km\;s$^{-1}$.}
    \label{fig:GP_cornerRVjoint}
\end{figure*}

\begin{figure*}
	\includegraphics[width=\textwidth]{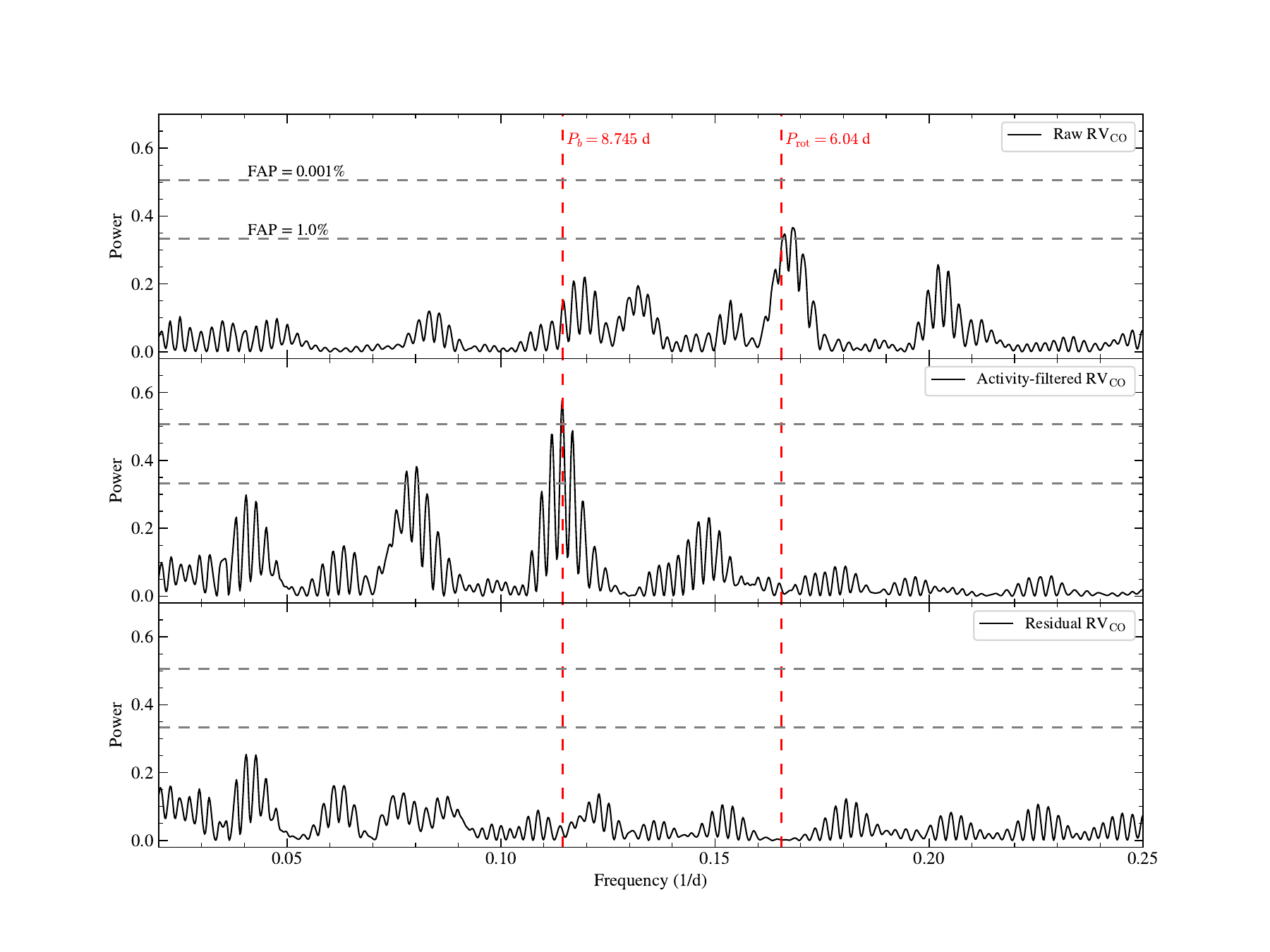}
    \caption{Generalised Lomb–Scargle periodogram computed for the RV data from CO lines. The top panel shows the raw RV data, the middle panel the activity-filtered RV, and the bottom panel the residual RV (i.e., activity- and planetary-filtered RV). The vertical red lines correspond to \prot and $P_b$, while the horizontal dashed lines illustrate $1\%$ and $0.1\%$ FAP levels. Note that in this figure activity and planet signals are filtered using the solution from the joint (atomic + CO lines) GP model with a planet in a circular orbit.} 
    \label{fig:PER_RVCO_Planet}
\end{figure*}

In Sec.~\ref{sec:GPrv}, we discussed the joint modelling of the RV data from atomic and CO lines. Fig.~\ref{fig:GP_cornerRVjoint} presents the posterior distributions obtained from the MCMC exploration of the 8-parameters allowed to vary in the GP+Planet model. The periodogram of the GP model with a planet is shown in Fig.~\ref{fig:PER_RVCO_Planet} for the CO line data. We observe a periodic signal at $8.745$\;days in the activity-filtered RV data (middle panel) and, once the planetary signal is removed (bottom panel), no periodicity remains in the residual data.

\begin{figure*}
	\includegraphics[width=0.9\textwidth]{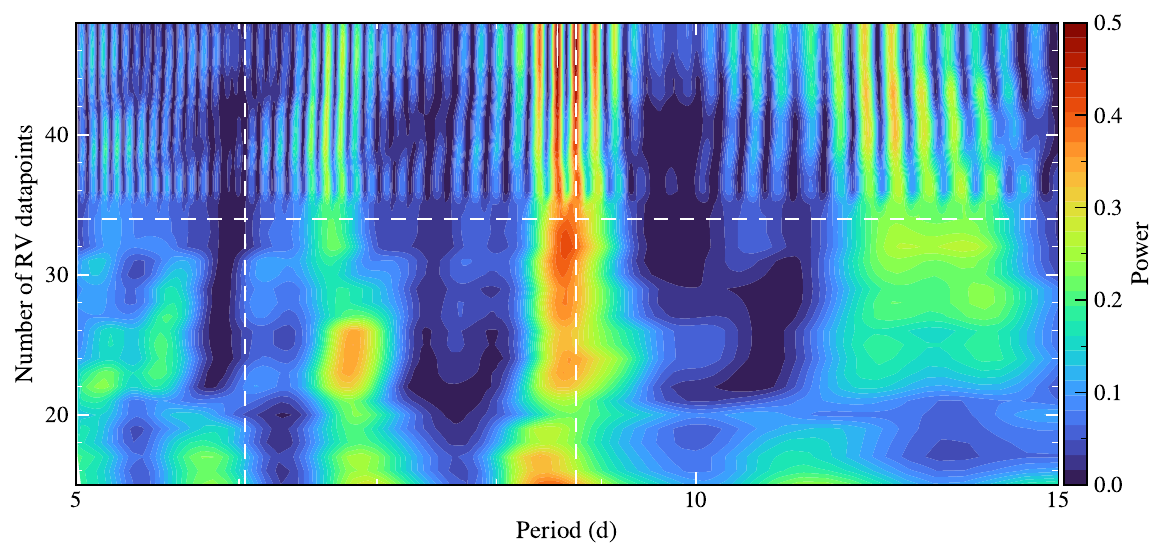}
    \caption{Stacked periodogram of the activity-filtered RV data from CO lines of GM Aur. The vertical dashed lines highlight \prot and $P_b$, while the horizontal line marks the transition to the last observing season (starting at December 1, 2022). The color scale depicts the power in the periodogram.}
    \label{fig:STACKEDPER_RVCO_Planet}
\end{figure*}

As discussed by \citet{MCC17}, the detection of planet-induced RV signals should increase its significance with the number of observations, contrary to activity signals that are quasi-periodic in nature, as planets induce a coherent signal stable over time. Fig.~\ref{fig:STACKEDPER_RVCO_Planet} illustrates how the power of the periodic signal at $8.745$\;d increases with the number of observations, corroborating the planet detection. We can also note a strong peak at $8.54$\;d, which corresponds to the 1-year alias of the true orbital period of $8.745$\;d.

\subsection{Independent fit of the RV data from atomic and CO lines}
\begin{figure*}
	\includegraphics[width=\textwidth]{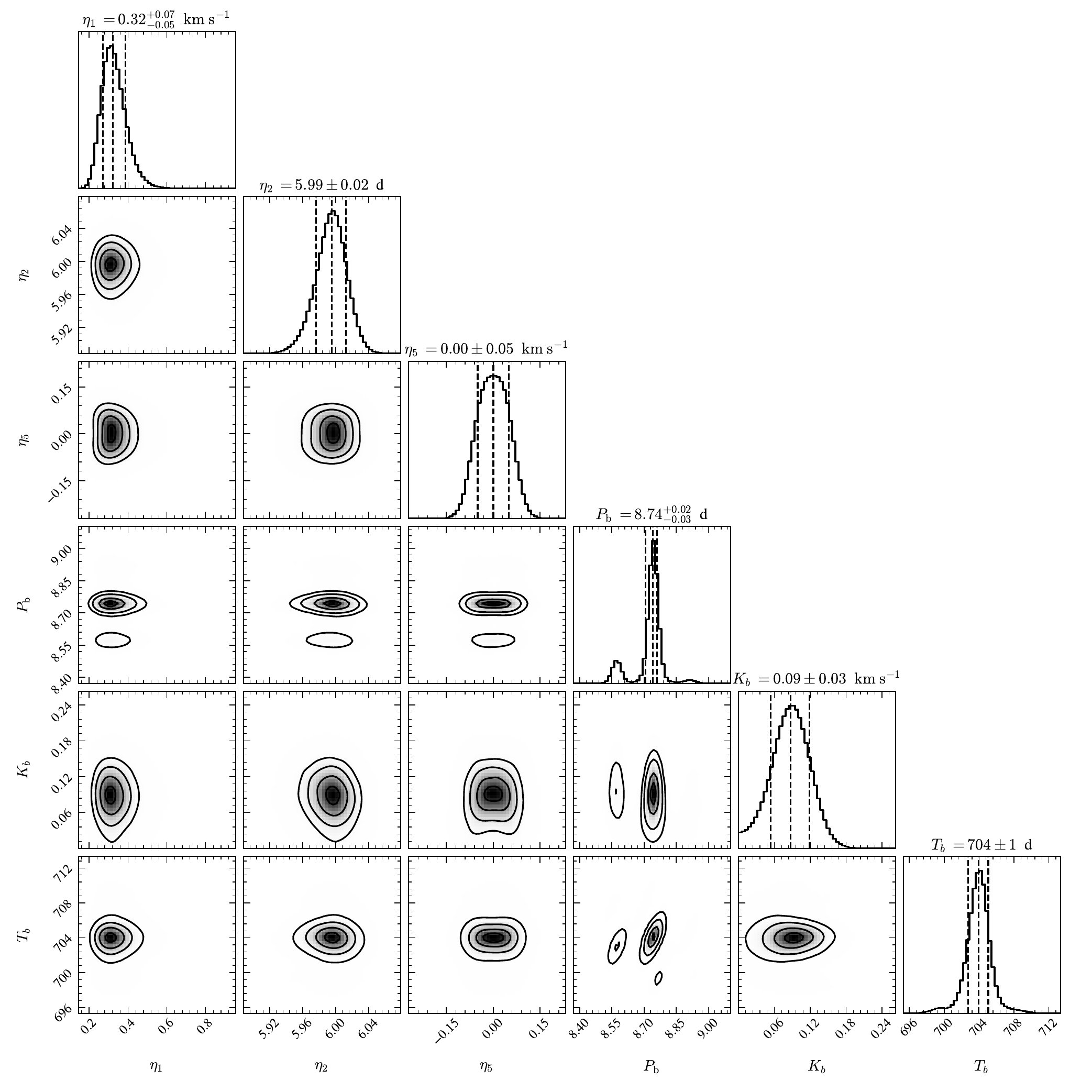}
    \caption{Similar to Fig.~\ref{fig:GP_cornerRVjoint}, but for GP+Planet model of the RV data from atomic lines.}
    \label{fig:GP_cornerRV}
\end{figure*}

\begin{figure*}
	\includegraphics[width=\textwidth]{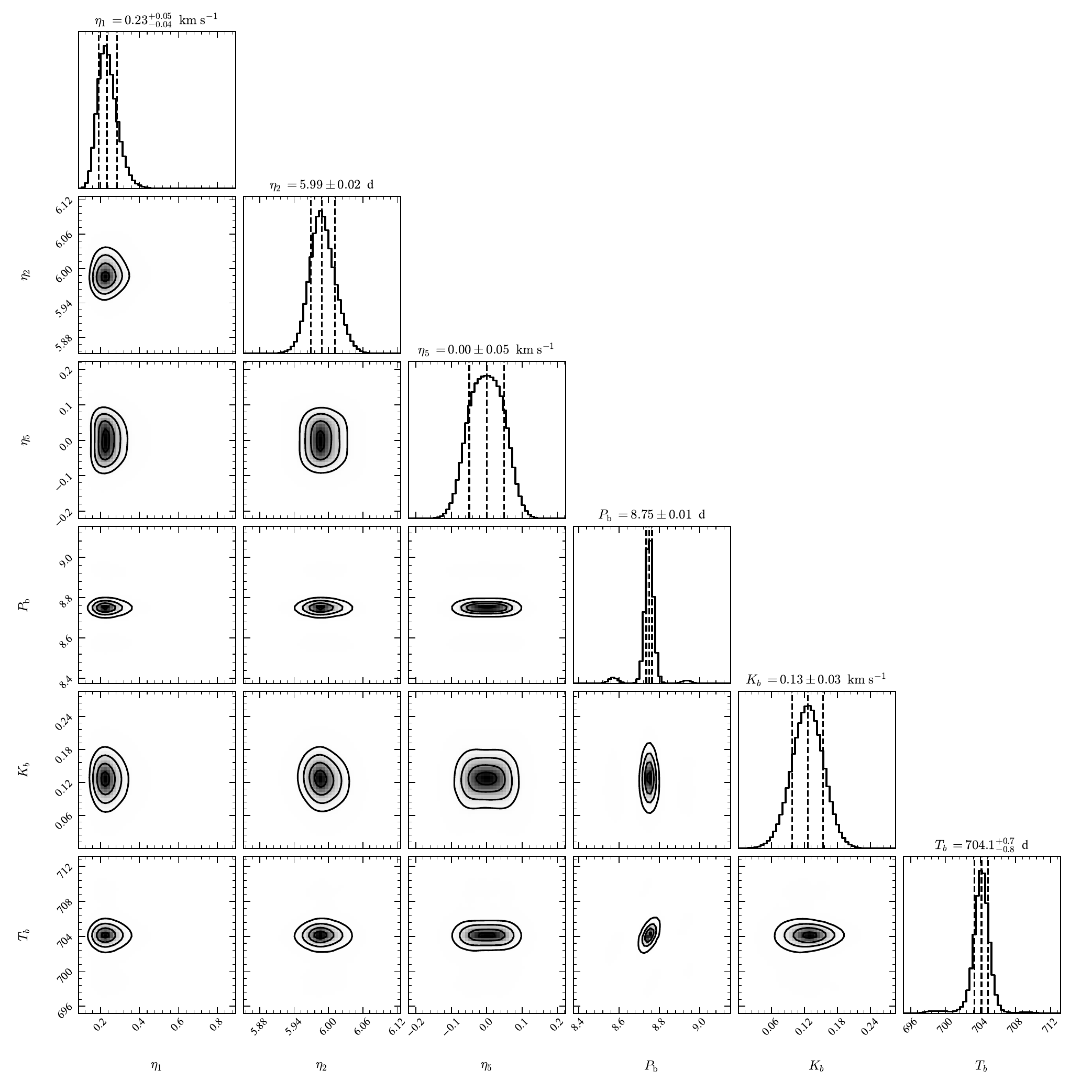}
    \caption{Similar to Fig.~\ref{fig:GP_cornerRVjoint}, but for GP+Planet model of the RV data from CO bandhead lines.}
    \label{fig:GP_cornerRVCO}
\end{figure*}

In this section, we illustrate the independent GP regression ran on the RV data from atomic and CO lines. The posterior distributions obtained from the MCMC search are shown in Figs.~\ref{fig:GP_cornerRV} and \ref{fig:GP_cornerRVCO}. The planetary solution obtained with the RV data from atomic lines agrees within the error bar with that from CO lines. While the former is detected at a $3\sigma$ level, the latter is detected at a $4\sigma$ level (see Table~\ref{tab:gpRVatomicfitparams}). Both results are in agreement with the planetary solution found in the joint fit described in the previous section.

\bsp	
\label{lastpage}
\end{document}